\documentclass[12pt]{article}
\usepackage{amsmath}
\usepackage{amssymb}
\usepackage{graphicx}
\usepackage{ifthen}
\usepackage{verbatim}
\usepackage{psfrag}

\newlength{\captionwidth}
\newsavebox{\tempbox}
\newcommand{\mycaption}[2]{%
\par\vspace{10pt}\sbox{\tempbox}{Figure #1: #2}%
\ifthenelse{\lengthtest{\wd\tempbox>\captionwidth}}%
{\sbox{\tempbox}{Figure.#1:\ }%
\addtolength{\captionwidth}{-\wd\tempbox}%
\mbox{Figure #1:\ }\parbox[t]{\captionwidth}{\small\textit{#2}}}%
{Figure #1: {\small\textit{#2}}}}%

\numberwithin{equation}{section}

\allowdisplaybreaks
\begin{document}
\thispagestyle{empty}
\begin{flushright}
\texttt{hep-th/0505083}\\
\texttt{OU-HET 528}\\
May 2005
\end{flushright}
\bigskip
\bigskip
\begin{center}
{\Large \textbf{Gravitational Quantum Foam}}\\   
{\Large \textbf{and Supersymmetric Gauge Theories}}
\end{center}
\bigskip
\bigskip
\renewcommand{\thefootnote}{\fnsymbol{footnote}}
\begin{center}
Takashi Maeda
\footnote{E-mail: \texttt{maeda@het.phys.sci.osaka-u.ac.jp}}, 
Toshio Nakatsu
\footnote{E-mail: \texttt{nakatsu@het.phys.sci.osaka-u.ac.jp}}, 
Yui Noma
\footnote{E-mail: \texttt{yuhii@het.phys.sci.osaka-u.ac.jp}}
and 
Takeshi Tamakoshi
\footnote{E-mail: \texttt{tamakoshi@het.phys.sci.osaka-u.ac.jp}}\\
\bigskip
{\small 
\textit{Department of Physics, Graduate School of Science, 
Osaka University,\\
Toyonaka, Osaka 560-0043, Japan\\}}
\end{center}
\bigskip
\bigskip
\renewcommand{\thefootnote}{\arabic{footnote}}
\begin{abstract}
We study K$\ddot{\mbox{a}}$hler gravity on local 
$SU(N)$ geometry and describe precise correspondence 
with certain supersymmetric gauge theories and random 
plane partitions. 
The local geometry is discretized, 
via the geometric quantization, 
to a foam of an infinite number of gravitational quanta. 
We count these quanta in a relative manner by 
measuring a deviation of the local geometry 
from a singular Calabi-Yau threefold, 
that is a $A_{N-1}$ singularity fibred over $\mathbb{P}^1$.  
With such a regularization prescription, 
the number of the gravitational quanta becomes finite 
and turns to be the perturbative prepotential for 
five-dimensional $\mathcal{N}=1$ supersymmetric 
$SU(N)$ Yang-Mills.  
These quanta are labelled by lattice points 
in a certain convex polyhedron on $\mathbb{R}^3$. 
The polyhedron becomes obtainable from 
a plane partition which is the ground state 
of a statistical model of random plane partition 
that describes the exact partition function for the gauge theory. 
Each gravitational quantum of the local geometry 
is shown to consist of $N$ unit cubes of plane partitions.

\end{abstract}

\setcounter{footnote}{0}
\newpage

\section{Introduction}

A surprising interpretation of a possible connection 
\cite{Crystal} between topological strings on Calabi-Yau threefolds 
and statistical models of crystal melting, 
known as random plane partitions,  has been advocated. 
It is conjectured \cite{quantum foam} that  
the statistical model is nothing but 
a quantum gravitational path integral involving 
fluctuations of K$\ddot{\mbox{a}}$hler geometry and topology,  
by interpreting plane partitions as gravitational quantum foams 
in K$\ddot{\mbox{a}}$hler gravity \cite{kahler gravity}, 
that is the target space field theory of topological $A$-model 
strings \cite{topological A_model}. 
It is shown \cite{quantum foam} that 
the topological vertex counting \cite{topological vertex} 
of the $A$-model partition function 
on a local toric Calabi-Yau threefold, 
which involves from the worldsheet viewpoint 
sums over holomorphic maps to the target space,  
is reproduced from quantum fluctuations 
of K$\ddot{\mbox{a}}$hler gravity 
on the fixed macroscopic background 
by assuming that the fluctuations are quantized 
in the unit of the string coupling constant $g_{st}$.

The topological vertex countings  
are known \cite{Iqbal-Eguchi} 
to bring about the gauge instanton contributions 
to the exact partition functions \cite{N-O} 
for supersymmetric gauge theories. 
However, the exact partition functions themselves are 
fully reproduced \cite{MNTT} by random plane partitions. 
Thinking over the conjecture, 
this indicates that 
even background geometries 
in K$\ddot{\mbox{a}}$hler gravity 
also have a microscopic origin in the statistical models, 
given by certain plane partitions, 
and emerge from the quantum foams by taking the semi-classical 
limit $g_{st}\rightarrow 0$. 
Further developing consideration along this line 
might lead to an explanation of microscopic generation of 
the target spaces for perturbative string theories. 
From the gauge theory viewpoint, 
it could  deepen our understanding 
of the geometric engineering \cite{geometric engineering} 
and will provide another formulation for gauge/gravity 
correspondence in string theories.

In this article we take the first step towards that direction. 
We study K$\ddot{\mbox{a}}$hler 
gravity on local $SU(N)$ geometry 
and describe precise correspondence 
with certain supersymmetric gauge theories 
and random plane partitions. 
The local geometry is a noncompact toric Calabi-Yau threefold  
and is considered as an ALE space with $A_{N-1}$ singularity 
fibred over $\mathbb{P}^1$. 
When K$\ddot{\mbox{a}}$hler forms $\omega$ on algebraic varieties 
are quantized in the unit of $g_{st}$, 
one is naturally led to an idea of geometric quantization 
\cite{geometric quantization} of the varieties.  
The geometric quantization requires a holomorphic line bundle $L$ 
on a variety $X$ whose first Chern class is equal to 
the quantized K$\ddot{\mbox{a}}$hler class.   
Physical states of the theory are given by the global sections,  
and the physical Hilbert space $\mathcal{H}_{\omega}$ 
is identified with 
\begin{eqnarray}
\mathcal{H}_{\omega}=H^0(X,\,L)\,.
\label{eq:intro_1}
\end{eqnarray}

In the quantization, we regard the variety of dimension $n$, 
equipped with the quantized K$\ddot{\mbox{a}}$hler form, 
as a phase space. 
The least volume measurable quantum mechanically 
on this phase space is $\sim g_{st}^n$. 
The variety could be divided into pieces of the minimal volume, 
where each piece corresponds to a gravitational quantum,  
and thereby discretized to a foam of gravitational quanta.  
It is clear that the number of the quanta is given by 
$\mbox{dim}\, \mathcal{H}_{\omega}$ and 
estimated as  $\mbox{Vol}(X)/g_{st}^n$.

Thanks to that the local geometry is a toric variety, 
the gravitational quanta turn to be labelled 
by lattice points in a certain convex polyhedron 
on $\mathbb{R}^3$. 
Reflecting the noncompactness of the geometry, 
the polyhedron is unbounded and 
the number of the lattice points becomes infinite. 
Hence the physical Hilbert space is infinite dimensional. 
A convenient way to manipulate such an infinity is 
to count the quanta in a relative sense rather than 
in an absolute sense. 
We introduce the regularized 
dimensions by counting the deviation from a singular 
Calabi-Yau threefold, 
that is a $A_{N-1}$ singularity fibred over $\mathbb{P}^1$.

The regularized dimensions turns to be a perturbative 
prepotential for five-dimensional supersymmetric gauge  
theory at the semi-classical limit $g_{st}\rightarrow 0$. 
This is a manifestation of gauge/gravity correspondence.

A possible relation between the quantized $SU(N)$ geometry 
and five-dimensional $\mathcal{N}=1$ supersymmetric $SU(N)$ 
Yang-Mills is suggested in \cite{MNTT} through the study of 
a statistical model of random plane partitions. 
After translating the quantized K$\ddot{\mbox{a}}$hler 
parameters into the gauge theory parameters as in \cite{MNTT}, 
the regularized dimensions satisfies 
\begin{eqnarray}
g_{st} \cdot 
\mbox{dim}\, \mathcal{H}_{\omega}
=
-\frac{1}{\hbar^2} 
\Bigl\{
\mathcal{F}^{pert}_{\mbox{{\scriptsize 5dSYM}}}
+O(\hbar) 
\Bigr\}\,, 
\label{eq:intro_2}
\end{eqnarray}
where 
$\mathcal{F}^{pert}_{\mbox{{\scriptsize 5dSYM}}}$
is the perturbative prepotential 
for five-dimensional $\mathcal{N}=1$ supersymmetric 
$SU(N)$ Yang-Mills with the Chern-Simons term 
\cite{Seiberg,triple intersection}. 
The five-dimensional theory is living on 
$\mathbb{R}^4 \times S^1$, 
where the radius $R$ of the circle in the fifth dimension 
is large (the large radius limit).  
The parameter $\hbar$ plays the similar role as $g_{st}$ 
in the statistical model, 
and relates with the string coupling constant  
by $g_{st}=2R\hbar$. 
The Chern-Simons coupling constant is quantized 
in accord with the framing of the local geometry   
that labels possible fibrations over $\mathbb{P}^1$. 
The above might lead a rationale for 
the geometric engineering \cite{geometric engineering}, 
which originally dictates that four-dimensional $\mathcal{N}=2$ 
supersymmetric $SU(N)$ Yang-Mills is realized 
by the $SU(N)$ geometry.

It is shown in \cite{MNTT,MNTT2} that 
a certain model of random plane partitions gives 
the exact partition function \cite{N-O} 
for the five-dimensional Yang-Mills. 
It is often convenient to identify plane partitions 
with the three-dimensional Young diagrams. 
By regarding each cube as a lattice point, 
the three-dimensional diagrams can be interpreted  
as the sets of lattice points in $\mathbb{R}^3$.

The ground state of the statistical model is 
called ground plane partition. 
The ground state energy turns out to relate with 
the perturbative prepotential by  
\begin{eqnarray}
E_{min}
=\frac{1}{\hbar^2} 
\Bigl\{
\mathcal{F}^{pert}_{\mbox{{\scriptsize 5dSYM}}}
+O(\hbar) 
\Bigr\}\,. 
\label{eq:intro_3}
\end{eqnarray}
The ground plane partition is determined 
uniquely from a ground partition.  
It is well known that partitions are realized 
in the Fock representation of two-dimensional 
complex fermions. Ground partitions appear naturally 
in an alternative realization  of the Fock representation 
by exploiting $N$ component complex fermions and 
get connected to K$\ddot{\mbox{a}}$hler gravities 
on a $A_{N-1}$ singularity or the ALE space 
\cite{MNTT}.

The ground plane partition  
leads to the quantum foam of local $SU(N)$ geometry.
In counting the deviation 
from the singular Calabi-Yau threefold,  
one needs to introduce another convex polyhedron 
which includes the original one. 
The complement, 
that becomes a bounded three-dimensional solid, 
measures the deviation. 
This three-dimensional solid becomes obtainable   
from the ground plane partition at the semi-classical 
limit $\hbar \rightarrow 0$. 
The lattice which is used 
to label the gravitational quanta 
becomes a sublattice of degree $N$ of the lattice 
on which plane partitions are drawn. 
It follows that each gravitational quantum of 
the $SU(N)$ geometry consists of $N$ cubes 
of plane partitions.

This article is organized as follows;  
we begin Section 2 by a brief review on toric geometries,  
and give the toric description of local $SU(N)$ geometry. 
We study K$\ddot{\mbox{a}}$hler gravity 
on the local geometry in Section 3.  
The regularized dimensions of the physical Hilbert space 
is introduced by using algebraic toric geometry. 
The discretization of the local 
geometry into the foam of gravitational quanta is explained  
by communicating with the gauged linear $\sigma$-model 
description of the geometry.  
We also argue the validity of 
the regularized dimensions by showing how the definition fits to 
the Riemann-Roch formula. 
In Section 4 we examine the statistical model of random 
plane partitions and show how the K$\ddot{\mbox{a}}$hler 
gravity is seen by the random plane partitions. 
In Appendix we summarize two-dimensional complex fermions 
which are used in the last section.

\section{Local $SU(N)$ Geometry as Toric Variety}

We start with a brief review on toric varieties.  
Further information on algebraic toric geometries  
can be found in \cite{Fulton} and \cite{Oda}. 
We then provide a toric description of local $SU(N)$ geometry, 
which is a noncompact toric Calabi-Yau threefold.

\subsection{A review on toric varieties}
\label{review on toric geometry}

A toric variety $X$ is an algebraic variety that contains 
an algebraic torus $T$ as a dense open subset, equipped 
with an action of $T$ on $X$ which extends the natural action 
of $T$ on itself. 
An $n$-dimensional toric variety  
is constructed from a $\mathbb{Z}^n$-lattice $N$ 
and a fan $\Delta$, 
which is a collection of rational strongly convex polyhedral 
cones $\sigma$ in $N_{\mathbb{R}}=N \otimes \mathbb{R}$, 
satisfying the conditions;  
{\it i) every face of a cone in $\Delta$ is also a cone in $\Delta$,}  
and 
{\it ii) the intersection of two cones in $\Delta$ is a face of each.} 
A rational strongly convex polyhedral cone in $N_{\mathbb{R}}$ 
is a cone with apex at the origin, generated by a finite number 
of vectors in the lattice $N$. 
Such a cone will be called simply as a cone in $N$.

Let $M$ denote the dual lattice and 
$\langle \,,\, \rangle$ be the dual pairing. 
For a cone $\sigma$ in $N$, the dual cone $\sigma^{\vee}$ 
is the set of vectors in 
$M_{\mathbb{R}}=M \otimes \mathbb{R}$ 
which are nonnegative on $\sigma$. 
This gives a finitely generated commutative semigroup 
which is a sub-semigroup of the dual lattice. 
\begin{eqnarray}
\mathcal{S}_{\sigma}
=
M \cap \sigma^{\vee}
=
\Bigl \{ u \in M \,;\,
\langle u, v \rangle 
\geq 0 
\,\,\,
\mbox{for} 
\,\,\forall v \in \sigma 
\Bigr \}. 
\label{Ssigma}
\end{eqnarray}
The corresponding group ring 
$\mathbb{C}[\mathcal{S}_{\sigma}]$ 
is a finitely generated commutative algebra. 
As a complex vector space 
the group ring has a basis $\chi_m$, 
where $m$ runs over $\mathcal{S}_{\sigma}$, 
with multiplication $\chi_m \cdot \chi_{m'}=\chi_{m+m'}$. 
Generators for $\mathcal{S}_{\sigma}$ 
provide generators for $\mathbb{C}[\mathcal{S}_{\sigma}]$. 
This commutative algebra determines an affine variety 
$U_{\sigma}$ by 
\begin{eqnarray} 
U_{\sigma}=\mbox{Spec}(\mathbb{C}[\mathcal{S}_{\sigma}])\,.
\label{afine Spec}
\end{eqnarray} 
The points correspond to semigroup homomorphisms 
from $\mathcal{S}_{\sigma}$ to $\mathbb{C}$, 
where $\mathbb{C}$ is understood as 
an abelian semigroup by multiplication. 
Thus we can write 
\begin{eqnarray}
U_{\sigma}=
\mbox{Hom}_{sgr}
(\mathcal{S}_{\sigma},\mathbb{C}). 
\label{afiine variety}
\end{eqnarray}
For a semigroup homomorphism $\alpha$ from 
$\mathcal{S}_{\sigma}$ to $\mathbb{C}$ and 
$m \in \mathcal{S}_{\sigma}$, 
the value of $\chi_m$ at the point $\alpha$ is 
given by $\chi_m(\alpha)=\alpha(m)$.  
The algebraic torus is identified with  
$\mbox{Hom}_{gr}(M,\mathbb{C}^{\ast})$,     
and acts on the affine variety as 
\begin{eqnarray}
\begin{array}{ccc}
T  \times  U_{\sigma} & 
\longrightarrow & 
U_{\sigma}  \\[2mm]
(t,\alpha) & 
\longmapsto & 
t \circ \alpha\,,  
\end{array}
\label{torus action}
\end{eqnarray}
where 
$t \circ \alpha (m) \equiv 
t(m)\alpha(m)$ 
for all 
$m \in \mathcal{S}_{\sigma}$.

Let us look at a basic example. 
Let $e_1,\cdots,e_n$ be a basis for $N$ and 
$e_1^{*},\cdots,e_n^{*}$ be the dual basis for $M$. 
Let $\sigma$ be the $k$-dimensional cone with the 
generators $e_1,\cdots,e_k$. Then 
$\mathcal{S}_{\sigma}=
\mathbb{Z}_{\geq 0}e^*_1+\cdots+
\mathbb{Z}_{\geq 0}e^*_k+
\mathbb{Z}e^*_{k+1}+\cdots+ 
\mathbb{Z}e^*_n$. 
Write $X_i=\chi_{e_i^*}$. 
The group ring becomes 
$\mathbb{C}[\mathcal{S}_{\sigma}]=
\mathbb{C}
\bigl[X_i\,(1\leq i \leq k),\, 
X_j,X_j^{-1}\,
(k+1 \leq j \leq n)
\bigr]$. 
Therefore we obtain 
$U_{\sigma}=(\mathbb{C})^k \times (\mathbb{C}^{\ast})^{n-k}$. 
This shows that 
if $\sigma$ is generated by $k$ elements which can be completed 
to a basis for $N$, 
the corresponding affine toric variety $U_{\sigma}$ is 
a product 
$(\mathbb{C})^k \times (\mathbb{C}^{\ast})^{n-k}$. 
In particular, such affine toric varieties are nonsingular.

The toric variety $X_{\Delta}$ is obtained  
from a fan $\Delta$ by taking a disjoint union of 
the affine toric varieties $U_{\sigma}$ 
for each $\sigma \in \Delta$, and patching them together by 
using the identity 
$U_{\sigma} \cap U_{\tau}=U_{\sigma\, \cap\, \tau}$. 
\begin{eqnarray}
X_{\Delta}=
\bigcup_{\sigma \in \Delta}
U_{\sigma}. 
\label{toric variety}
\end{eqnarray}
The torus actions on the affine varieties 
are compatible with one another by the gluing,  
and give rise to the torus action on $X_{\Delta}$. 
The zero-dimensional cone $\{ 0 \}$ determines 
the affine variety $U_{\{0\}}=\mbox{Hom}_{gr}(M,\mathbb{C}^{\ast})$.  
It is nothing but the torus which becomes 
a dense open subset of $X_{\Delta}$.

It is possible to decompose a toric variety  
into a disjoint union of its orbits by the torus action. 
It appears one such orbit $O_{\tau}$ 
for each cone $\tau \in \Delta$. 
Let $\tau^{\perp}$ be the set of vectors in 
$M_{\mathbb{R}}$ which vanish on $\tau$. 
For any cone $\sigma$ that contains $\tau$ as a face, 
we consider the embedding 
\begin{eqnarray}
\mbox{Hom}_{sgr}
(\mathcal{S}_{\sigma}\cap \tau^{\perp},\mathbb{C}^{\ast})\, 
\hookrightarrow \,
U_{\sigma}
\end{eqnarray}
given by extension by zero on the complement 
$\mathcal{S}_{\sigma}\setminus 
(\mathcal{S}_{\sigma}\cap \tau^{\perp})$. 
By the embedding,  
$\mbox{Hom}_{sgr}
(\mathcal{S}_{\sigma}\cap \tau^{\perp},\mathbb{C}^{\ast})$ 
becomes an orbit of the torus action 
(\ref{torus action}). 
The strong convexity of $\sigma$, 
which is translated to the nondegeneracy 
$\mathcal{S}_{\sigma}+(-\mathcal{S}_{\sigma})=M$, makes  
the orbit identical to 
$\mbox{Hom}_{gr}(M \cap \tau^{\perp},\mathbb{C}^{\ast})$, 
which is $(\mathbb{C}^{\ast})^{n-k}$ if $\mbox{dim}\, \tau=k$. 
By letting 
\begin{eqnarray}
O_{\tau}
=\mbox{Hom}_{gr}(M \cap \tau^{\perp},\mathbb{C}^{\ast}), 
\label{O}
\end{eqnarray}
we see that each affine variety in (\ref{toric variety}) is a disjoint 
union of those orbits $O_{\tau}$ for which $\sigma$ contains 
$\tau$ as a face. 
\begin{eqnarray}
U_{\sigma}=
\coprod_{\tau < \sigma}O_{\tau}, 
\label{U by O}
\end{eqnarray}
where $\tau < \sigma$ means that $\tau$ is a face of $\sigma$. 
Therefore $X_{\Delta}$ becomes a disjoint union 
of the orbits $O_{\sigma}$ over $\sigma \in \Delta$. 
\begin{eqnarray}
X_{\Delta}=
\coprod_{\sigma \in \Delta}
O_{\sigma}\,. 
\label{X by O}
\end{eqnarray}

The closure of $O_{\sigma}$ becomes a closed subvariety 
of $X_{\Delta}$ which is again a toric variety. 
Let us denote the closure of $O_{\sigma}$ by 
$V(\sigma)$. 
This turns out to be a disjoint union of 
the orbits $O_{\tau}$ for which 
$\tau$ contains $\sigma$ as a face. 
\begin{eqnarray}
V(\sigma)=
\coprod_{\tau > \sigma}
O_{\tau}.
\label{V}
\end{eqnarray}

\subsection{Local $SU(N)$ geometry as toric variety}
\label{toric data}

We provide a toric description of local $SU(N)$ geometry.
Local $SU(N)$ geometry is an ALE space with 
$A_{N-1}$ singularity fibred over $\mathbb{P}^1$. 
The fibration over $\mathbb{P}^1$ is labeled by an integer 
$k_{\mbox{\tiny{C.S.}}} \in [0,\,N]$, which is called the framing.

\begin{figure}[htb]
 \psfrag{r0}{$\rho_0$}
 \psfrag{r1}{$\rho_1$}
 \psfrag{r2}{$\rho_2$}
 \psfrag{r3}{$\cdots$}
 \psfrag{r4}{$\rho_N$}
 \psfrag{rN+1}{$\rho_{N+1}$}
 \psfrag{rN+2}{$\rho_{N+2}$}
 \psfrag{s0}{$\sigma_{0}$}
 \psfrag{s1}{$\sigma_{1}$}
 \psfrag{s2}{$\sigma_{2}$}
 \psfrag{s3}{$\sigma_{3}$}
 \psfrag{s4}{}
 \psfrag{s5}{}
 \psfrag{s6}{$\sigma_{2N-2}$}
 \psfrag{s7}{$\sigma_{2N-1}$}
 \psfrag{o}{$0$}
 \begin{center}
  \includegraphics[scale=.8]{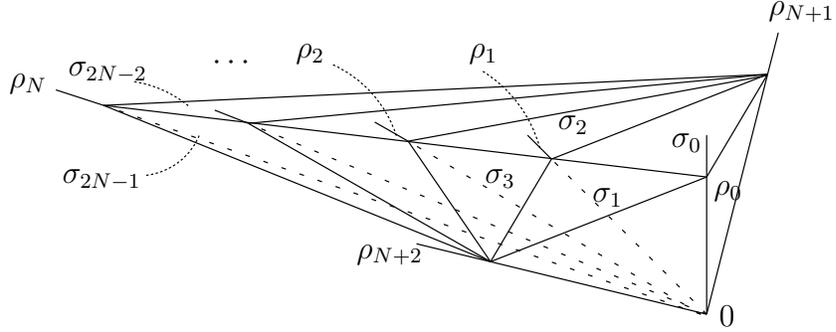}
  \caption{\textit{Cones relevant to describe the $SU(N)$ geometry.}}
  \label{fig;SU(N)fan}
 \end{center}
\end{figure}

Let $N$ be a three-dimensional lattice with generators 
$e_1,e_2$ and $e_3$.  
Let $v_i$ be the following $N+3$ elements in $N$.   
\begin{eqnarray}
v_i=ie_2+e_3    
\hspace{4mm} 
(0 \leq i \leq N), 
\hspace{6mm}
v_{N+1}=e_1+e_3, 
\hspace{6mm}
v_{N+2}=-e_1+\widetilde{k}_{\mbox{\tiny{C.S.}}}e_2+e_3,
\label{vi}
\end{eqnarray}
where $\widetilde{k}_{\mbox{\tiny{C.S.}}}\equiv N-k_{\mbox{\tiny{C.S.}}}$.
We introduce three-dimensional cones 
$\sigma_0,\sigma_1,\,\cdots\,,\sigma_{2N-1}$ 
in $N$ as follows.   
\begin{eqnarray}
\sigma_{2k}&=&
\mathbb{R}_{\geq 0}
v_k+
\mathbb{R}_{\geq 0}
v_{k+1}+
\mathbb{R}_{\geq 0}
v_{N+1}, 
\nonumber \\
\sigma_{2k+1}&=&
\mathbb{R}_{\geq 0}
v_k+
\mathbb{R}_{\geq 0}
v_{k+1}+
\mathbb{R}_{\geq 0}
v_{N+2}.
\label{3d cones}
\end{eqnarray}
All these cones and their faces 
constitute the fan $\Delta$ which is relevant to describe  
local $SU(N)$ geometry with the framing $k_{\mbox{\tiny{C.S.}}}$. 
See Figure \ref{fig;SU(N)fan}. 
Generators for any cone $\sigma \in \Delta$ are taken from 
the elements $v_i$.  
It is easy to see that these generators can be always completed 
to a basis for $N$. 
This shows that the corresponding affine variety $U_{\sigma}$  
is nonsingular. 
Therefore the corresponding toric variety is nonsingular. 

There exist $N$ linear relations 
among $v_i$ owing to the dimensionality.
\begin{eqnarray}
v_{i-1}-2v_i+v_{i+1}&=&0 
\hspace{8mm}
(1 \leq i \leq N-1)\,, 
\nonumber \\ 
(\widetilde{k}_{\mbox{\tiny{C.S.}}}-2)v_0 
-\widetilde{k}_{\mbox{\tiny{C.S.}}}v_1+v_{N+1}+v_{N+2}
 &=&0\,. 
\label{relations among vi}
\end{eqnarray}
These relations are conveniently expressed 
in the matrix form by  
\begin{eqnarray}
\bigl( \hspace{1mm} 
v_0,\cdots,v_i,\cdots,v_{N+2}
\hspace{1mm}
\bigr)\, 
Q=0\,,
\label{relations by Q}
\end{eqnarray}
where
\begin{eqnarray} 
Q=
\bigl(
Q_{i a}
\bigr)_{0\leq i \leq N+2,\,1 \leq a \leq N}
=
\left( 
\begin{array}{cccccc}
1  &    &        &    & \widetilde{k}_{\mbox{\tiny{C.S.}}}-2 \\
-2 & 1  &        &    &  -\widetilde{k}_{\mbox{\tiny{C.S.}}}  \\
1  & -2 & \ddots &    &    \\
   &  1 & \ddots & 1  &    \\
   &    & \ddots & -2 & 0  \\  
   &    &        & 1  & 0  \\
   &    &        &    & 1  \\
   &    &        &    & 1
\end{array}
\right)\,. 
\label{Q}
\end{eqnarray}
The matrix $Q$ is identified \cite{Mirror} 
with a charge matrix which appears in 
the gauged linear $\sigma$-model description of 
local $SU(N)$ geometry.

For the later convenience 
we name the lower dimensional cones of $\Delta$.
There appear one-dimensional cones 
$\rho_0,\rho_1,\,\cdots\,,\rho_{N+2}$,  
and two-dimensional cones 
$\gamma_0,\gamma_1,\,\cdots\,,\gamma_{3N+1}$. 
See Figure \ref{fig;SU(N)fan}. 
They are given by    
\begin{eqnarray}
\rho_i
&=&
\mathbb{R}_{\geq 0}v_i\,,
\label{1d cones}
\\[2mm]
\gamma_{3k}
&=&
\mathbb{R}_{\geq 0}v_k+
\mathbb{R}_{\geq 0}v_{N+1}\,,
\nonumber \\
\gamma_{3k+1}
&=&
\mathbb{R}_{\geq 0}v_k+
\mathbb{R}_{\geq 0}v_{N+2}\,, 
\label{2d cones} 
\\
\gamma_{3k+2}
&=&
\mathbb{R}_{\geq 0}v_k+
\mathbb{R}_{\geq 0}v_{k+1}\,.
\nonumber 
\end{eqnarray}

\section{K$\ddot{\mbox{a}}$hler Gravity 
on Local $SU(N)$ Geometry}

One obtains 
local $SU(N)$ geometry from the fan 
$\Delta 
=\bigl\{ \sigma_i,\,\gamma_j,\,\rho_k,\, \{0\} \bigr\}$ 
as a nonsingular toric variety $X_{\Delta}$. 
The two-dimensional cones $\gamma_j$ determine 
two-cycles $V(\gamma_j)$. 
Among them we can choose 
$V(\gamma_2)$ and $V(\gamma_{3k})$ with $1 \leq k \leq N-1$, 
as a basis for $H_2(X_{\Delta},\mathbb{Z})$. 
Vanishing cycles of the fibred $A_{N-1}$ singularity are 
$V(\gamma_{3k})$.  
The base $\mathbb{P}^1$ is identified with 
$V(\gamma_2)$ for the case of $k_{\mbox{\tiny{C.S.}}}=N$.   
One needs a slight modification for other framings.

Let $\omega$ be a K$\ddot{\mbox{a}}$hler two-form. 
The K$\ddot{\mbox{a}}$hler class 
$[\omega] \in H^2(X_{\Delta},\mathbb{R})$ 
is specified by the K$\ddot{\mbox{a}}$hler volumes 
of the two-cycles. 
\begin{eqnarray}
t_k \equiv \int_{V(\gamma_{3k})}
\!\!\! \omega\,, 
\hspace{8mm}
t_N \equiv \int_{V(\gamma_{2})}
\!\!\! \omega\,, 
\label{classical kahler parameters}
\end{eqnarray}
where the classical parameters $t$ 
take nonnegative real numbers. 
In what follows, 
as suggested in \cite{quantum foam}, 
we assume that 
K$\ddot{\mbox{a}}$hler forms are  
quantized in the unit of $g_{st}$.  
\begin{eqnarray}
\frac{1}{g_{st}}\bigl[\omega \bigr] 
\, \in 
H^2(X_{\Delta},\mathbb{Z})\,. 
\label{quantized kahler form}
\end{eqnarray}
They are specified by the quantized parameters 
$T$ which take nonnegative integers. 
\begin{eqnarray}
T_k
\equiv 
\frac{1}{g_{st}}
\int_{V(\gamma_{3k})}
\!\!\! \omega
\hspace{-3mm}
&&
\in \mathbb{Z}_{\geq 0} 
\hspace{5mm}
(k=1,2,\,\cdots\,,N-1)\,, 
\nonumber \\
T_{N}
\equiv
\frac{1}{g_{st}}
\int_{V(\gamma_{2})}
\!\!\! \omega 
\hspace{-3mm}
&& 
\in \mathbb{Z}_{\geq 0}
\,.
\label{quantized kahler parameters}
\end{eqnarray}

When the K$\ddot{\mbox{a}}$hler forms on algebraic varieties  
are quantized,  one is naturally led to an idea of geometric quantization 
of the varieties. The power of geometric quantizations for gravity theories 
may be seen in three-dimensional Chern-Simons gravity 
\cite{CS gravity} as investigated in \cite{coadjoint orbits}.

\subsection{Geometric quantization of local $SU(N)$ geometry}
\label{geometric quantization}

The geometric quantization requires, 
first of all, a holomorphic line bundle $L$
whose first Chern class is equal to 
the quantized K$\ddot{\mbox{a}}$hler class.   
\begin{eqnarray}
c_1(L)=\frac{1}{g_{st}}\bigl[\omega \bigr]\,. 
\label{BS condition}
\end{eqnarray}
Let $Pic(X)$ be the group of all line 
bundles on an algebraic variety $X$, 
modulo isomorphism. 
For each $[L] \in Pic(X)$ 
we can associate the first Chern class of the line bundle. 
This gives a map from $Pic(X)$ to $H^2(X,\mathbb{Z})$.  
In the present toric geometry 
the map turns to be bijective, 
and $Pic(X_{\Delta})$ becomes isomorphic 
to $H^2(X_{\Delta},\mathbb{Z})$. 
\begin{eqnarray}
\begin{array}{ccc}
Pic(X_{\Delta}) & 
\stackrel{\simeq}{\longrightarrow} & 
H^2(X_{\Delta},\mathbb{Z}) \\[2mm]
[L] & \longmapsto & c_1(L) 
\end{array} 
\label{Pic}
\end{eqnarray}
This shows that we can always find out 
the holomorphic line bundle $L$ 
with $c_1(L)=[\omega]/g_{st}$.

The condition (\ref{quantized kahler parameters}), 
which is interpreted 
as the Bohr-Sommerfeld quantization rule, 
is now translated to the following condition on the line bundle. 
\begin{eqnarray}  
\int_{V(\gamma_{3k})} 
\!\!\! c_1(L)
&\geq& 0
\hspace{8mm}
(k=1,2,\cdots \,,N-1)\,, 
\nonumber \\
\int_{V(\gamma_{2})}
\!\!\! c_1(L)
&\geq&  0\,. 
\label{ampleness 1}
\end{eqnarray}
This means that $L$ is an ample line bundle or 
generated by its sections. It should be noted that 
the ampleness implies \cite{Fulton} 
the vanishing of all higher sheaf cohomology groups of $L$. 
\begin{eqnarray}
H^p(X_{\Delta},L)=0 
\hspace{5mm} 
\mbox{for}
\hspace{3mm}
\forall p >0. 
\label{vanishing theorem}
\end{eqnarray}

Let $L$ be the holomorphic line bundle with 
$c_1(L)=[\omega]/g_{st}$. 
In the geometric quantization, 
physical states, more precisely, 
wave functions of physical states are prescribed 
as the holomorphic sections.  
These are vectors in 
$\Gamma(X_{\Delta},L)=H^0(X_{\Delta},L)$. 
The physical Hilbert space $\mathcal{H}_{\omega}$ becomes 
\begin{eqnarray}
\mathcal{H}_{\omega}=H^0(X_{\Delta},L)\,.
\label{def physical Hilbert space}
\end{eqnarray}
It will be seen subsequently that the holomorphic 
sections are labeled by lattice points in a convex 
polyhedron.

\subsection{Divisors and line bundles}
\label{divisor and line bundle}

On a toric variety,  
a Weil divisor is a finite formal sum 
of irreducible closed subvarieties 
of codimension one that are invariant under the torus action.  
These invariant subvarieties of codimension one 
correspond to rays or edges of the fan. 
In the $SU(N)$ geometry,  
the rays are named $\rho_i$ in (\ref{1d cones}) 
and the corresponding divisors are 
the closures of $O_{\rho_i}$. 
\begin{eqnarray}
D_i \equiv V(\rho_i), 
\label{divisors}
\end{eqnarray}
where $0 \leq i \leq N+2$.  
It follows from (\ref{V}) that compact divisors 
are $D_1,D_2,\cdots,D_{N-1}$.  
These are isomorphic to the Hirzebruch surfaces.

Weil divisors, 
which are the sums $\sum_i d_i D_i$ for integers $d_i$, 
determine holomorphic line bundles on the $SU(N)$ geometry, 
modulo linear equivalence, 
and vice versa. To see this, it is convenient 
to use Cartier divisors instead of Weil divisors. 
Let $D=\sum_{i=0}^{N+2}d_iD_i$ be a Weil divisor. 
The corresponding Cartier divisor is defined by specifying 
an element $l(\sigma)$ in $M/(M\cap \sigma^{\perp})$ 
for each $\sigma \in \Delta$. The element $l(\sigma)$ 
is determined by the following condition.
\begin{eqnarray}
\langle l(\sigma)\,,v_i \rangle=-d_i 
\hspace{7mm} 
\mbox{for}
\hspace{3mm}
\forall 
\rho_i < \sigma. 
\label{Cartier divisor}
\end{eqnarray}
In particular,  
for a maximal dimensional cone $\sigma$, 
this fixes $l(\sigma)$ as an element in $M$. 
Let $\mathcal{O}(D)$ denote the line bundle 
which is determined by $D$. 
It is described by specifying holomorphic sections 
on each affine variety $U_{\sigma}$ in $X_{\Delta}$. 
They are defined by 
\begin{eqnarray} 
\Gamma(U_{\sigma},\mathcal{O}(D))
&\equiv&
\mathbb{C}[\mathcal{S}_{\sigma}] 
\cdot 
\chi_{l(\sigma)} 
\nonumber \\
&=&
\bigoplus_{m \in \mathcal{S}_{\sigma}}
\mathbb{C}\chi_{m+l(\sigma)}\,.
\label{O(D)}
\end{eqnarray}

The global sections of the line bundle 
are sections which are common to 
$\Gamma(U_{\sigma},\mathcal{O}(D))$ for 
all $\sigma \in \Delta$. 
\begin{eqnarray}
\Gamma(X_{\Delta},\mathcal{O}(D))=
\bigcap_{\sigma \in \Delta} 
\Gamma(U_{\sigma},\mathcal{O}(D))\,. 
\label{def of global sections}
\end{eqnarray}
Let $\chi_m$ be a global section. 
It follows from (\ref{O(D)}) that 
$m$ is an element in $M$ which satisfies 
$m \in \mathcal{S}_{\sigma}+l_{\sigma}$ 
for any cone $\sigma$. Taking account of 
(\ref{Cartier divisor}), this is translated to 
the condition. 
\begin{eqnarray}
\langle\, m, v_i \,\rangle 
+d_i \geq 0 
\hspace{6mm} 
\mbox{for}
\hspace{2mm}
\forall 
\rho_i\,. 
\label{global section}
\end{eqnarray}

It is also helpful to provide the standard description 
via local trivialization. The line bundle is trivialized to 
$\mathcal{O}(D)
=\cup_{\sigma \in \Delta} (U_{\sigma} \times \mathbb{C})$. 
These trivializations are pieced together along their overlaps 
by using transition functions $g_{\tau \sigma}$. 
They are nowhere vanishing holomorphic functions 
on overlaps $U_{\sigma\, \cap\, \tau}=U_{\sigma}\cap U_{\tau}$. 
As can be read from (\ref{O(D)}), 
the transition functions are taken as   
\begin{eqnarray}
g_{\tau \sigma}=\chi_{l(\sigma)-l(\tau)}\,.
\label{transition function}
\end{eqnarray}
Since $\sigma \cap \tau$ is a face of both $\sigma$ and $\tau$, 
it follows from (\ref{Cartier divisor}) that $l(\sigma)-l(\tau)$ 
vanishes on $\sigma \cap \tau$. 
This means that $\chi_{l(\sigma)-l(\tau)}$ is actually 
a nowhere vanishing function on $U_{\sigma\, \cap\, \tau}$.

The transition functions (\ref{transition function}) 
determine the first Chern class of 
$\mathcal{O}(D)$, as a cocycle. 
Let us use an abbreviation 
$M(\sigma)\equiv M\cap \sigma^{\perp}$. 
The second cohomolgy group of $X_{\Delta}$ 
can be realized \cite{Fulton} as 
\begin{eqnarray}
H^2(X_{\Delta},\mathbb{Z})=
\mbox{Ker}
\Bigl( 
\bigoplus_{i<j}M(\sigma_i \cap \sigma_j)
\rightarrow 
\bigoplus_{i<j<k}
M(\sigma_i \cap \sigma_j \cap \sigma_k)
\Bigr)\,, 
\label{2nd cohomology}
\end{eqnarray}
where $\sigma_i$ are the three-dimensional cones. 
Thus, the first Chern class of $\mathcal{O}(D)$ 
is described by
\begin{eqnarray}
c_1(\mathcal{O}(D))
=\Bigl\{ 
l(\sigma_i)-l(\sigma_j) 
\Bigr\}\,.
\label{c1 of O(D)}
\end{eqnarray}

Let us compute the first Chern numbers of 
$\mathcal{O}(D)$. 
Let $e_1^*,e_2^*$ and $e_3^*$ be the dual basis for $M$. 
Taking account of (\ref{vi}) and (\ref{3d cones}), 
the corresponding Cartier divisor is determined by 
(\ref{Cartier divisor}) as follows.
\begin{eqnarray}
l(\sigma_{2k})&=& 
\Bigl\{ (k+1)d_k-kd_{k+1}-d_{N+1}\Bigr\}e_1^*
+
(d_k-d_{k+1})e_2^* 
\nonumber \\
&&
+\Bigl\{-(k+1)d_k+kd_{k+1}\Bigr\}e_3^*\,, 
\nonumber \\
l(\sigma_{2k+1})
&=& 
\Bigl\{(\widetilde{k}_{\mbox{\tiny{C.S.}}}-k-1)d_k
+(-\widetilde{k}_{\mbox{\tiny{C.S.}}}+k)d_{k+1}+d_{N+2}\Bigr\}e_1^*
+
(d_k-d_{k+1})e_2^* \nonumber \\
&&
+
\Bigl\{-(k+1)d_k+kd_{k+1}\Bigr\}e_3^*\,.  
\label{divisor for 3d cones for O(D)}
\end{eqnarray}
These give the cocycle as  
\begin{eqnarray}
l(\sigma_{2k})-l(\sigma_{2k-2})
&=&
(d_{k-1}-2d_k+d_{k+1})
(-ke_1^*-e_2^*+ke_3^*) 
\hspace{2mm}
\in M(\gamma_{3k})\,,
\nonumber \\
l(\sigma_{2k+1})-l(\sigma_{2k-1})
&=&
(d_{k-1}-2d_k+d_{k+1})
\Bigl\{
(-\widetilde{k}_{\mbox{\tiny{C.S.}}}+k)e_1^*-e_2^*+ke_3^*
\Bigr\} 
\hspace{2mm}
\in M(\gamma_{3k+1})\,, 
\nonumber \\
l(\sigma_1)-l(\sigma_0)
&=&
\Bigl\{
(\widetilde{k}_{\mbox{\tiny{C.S.}}}-2)d_0
-\widetilde{k}_{\mbox{\tiny{C.S.}}}d_1
+d_{N+1}+d_{N+2}
\Bigr\}e_1^* 
\hspace{2mm}
\in 
M(\gamma_2)\,. 
\label{cocycle of c1 for O(D)}
\end{eqnarray}
The first Chern numbers are read from 
(\ref{cocycle of c1 for O(D)}) as 
\begin{eqnarray}
\int_{V(\gamma_{3k})}c_1(\mathcal{O}(D))
&=&
d_{k-1}-2d_k+d_{k+1}\,,
\nonumber \\
\int_{V(\gamma_2)}c_1(\mathcal{O}(D))
&=&
(\widetilde{k}_{\mbox{\tiny{C.S.}}}-2)d_0
-\widetilde{k}_{\mbox{\tiny{C.S.}}}d_1
+d_{N+1}+d_{N+2}\,. 
\label{c1 numbers of O(D)}
\end{eqnarray}
Note that these are also expressed by using the charge 
matrix as 
\begin{eqnarray}
\Bigl(\,\, 
\int_{V(\gamma_{3k})}c_1(\mathcal{O}(D))\,\,, 
\int_{V(\gamma_2)}c_1(\mathcal{O}(D))\,\,
\Bigr)
= 
\bigl(
\,d_0, \cdots, d_{N+2}\, 
\bigr) 
Q\,. 
\label{c1 numbers by Q}
\end{eqnarray}

\subsection{Gauge$\slash$gravity correspondence 
for $k_{\mbox{\tiny{C.S.}}}=N$}
\label{gauge/gravity correspondence 1}

We first consider the geometry with $k_{\mbox{\tiny{C.S.}}}=N$. 
Cases of other framings will be treated in section 
\ref{gauge/gravity correspondence 2}. 
Let $\omega$ be a quantized K$\ddot{\mbox{a}}$hler form 
of the $SU(N)$ geometry. 
Let $\mathcal{O}(D)$ be the line bundle the first Chern class 
of which is the quantized K$\ddot{\mbox{a}}$hler class. 
\begin{eqnarray}
c_1(\mathcal{O}(D))
=\frac{1}{g_{st}}
\bigl[ \omega \bigr]\,. 
\label{BS condition for O(D)}
\end{eqnarray}
The Weil divisor $D=\sum_{i=0}^{N+2}d_iD_i$ is unique,  
up to linear equivalence. It follows from 
(\ref{quantized kahler parameters}) and 
(\ref{c1 numbers of O(D)}) that the quantized 
K$\ddot{\mbox{a}}$hler parameters are written by 
means of $d_i$ as 
\begin{eqnarray}
T_k 
&=& 
d_{k-1}-2d_k+d_{k+1} 
\hspace{8mm}
(k=1,2,\,\cdots\,,N-1)\,, 
\nonumber \\
T_N
&=& 
d_{N+1}-2d_0+d_{N+2}\,.
\label{Ti by di}
\end{eqnarray}

The divisor $D$ determines a rational convex polyhedron 
in $M_{\mathbb{R}}$ defined by 
\begin{eqnarray}
\mathcal{P}(D)\equiv 
\Bigl \{
x \in M_{\mathbb{R}}\,;\, 
\langle\, x,\, v_i \,\rangle 
+d_i \geq 0 
\hspace{3mm} 
\mbox{for} 
\hspace{2mm}
\forall \rho_i
\Bigr \}\, . 
\label{PD}
\end{eqnarray}
The global sections of $\mathcal{O}(D)$ become  
the physical Hilbert space $\mathcal{H}_{\omega}$. 
It follows from (\ref{global section}) that 
these are labeled by elements in $\mathcal{P}(D) \cap M$. 
Therefore we obtain 
\begin{eqnarray}
\mathcal{H}_{\omega}=
\bigoplus_{m\,\in\,\mathcal{P}(D)\,\cap\, M}
\mathbb{C} \chi_{m}\,. 
\label{physical Hilbert space}
\end{eqnarray}
\begin{figure}[htb]
 \psfrag{0}{$l(\sigma_0)$}
 \psfrag{1}{\hspace{-1mm}$l(\sigma_1)$}
 \psfrag{2}{$l(\sigma_2)$}
 \psfrag{3}{\hspace{-1mm}$l(\sigma_3)$}
 \psfrag{4}{\hspace{4mm}\rotatebox{45}{$\cdots$}}
 \psfrag{5}{\hspace{0mm}\rotatebox{90}{$\cdots$}}
 \psfrag{6}{$l(\sigma_{2N-4})$}
 \psfrag{7}{\hspace{-13mm}$l(\sigma_{2N-3})$}
 \psfrag{8}{$l(\sigma_{2N-2})$}
 \psfrag{9}{\hspace{-7mm}$l(\sigma_{2N-1})$}
 \psfrag{k0}{$K_0$}
 \psfrag{k1}{$K_1$}
 \psfrag{k2}{$K_2$}
 \psfrag{k3}{\rotatebox{60}{$\cdots$}}
 \psfrag{k4}{$K_{N-1}$}
 \psfrag{k5}{$K_{N}$}
 \psfrag{k6}{$K_{N+1}$}
 \psfrag{k7}{$K_{N+2}$}
 \psfrag{o1}{}
 \psfrag{o2}{}
 \psfrag{M}{\hspace{-2mm}$M_{\mathbb{R}}$}
 \begin{center}
  \includegraphics[scale=.8]{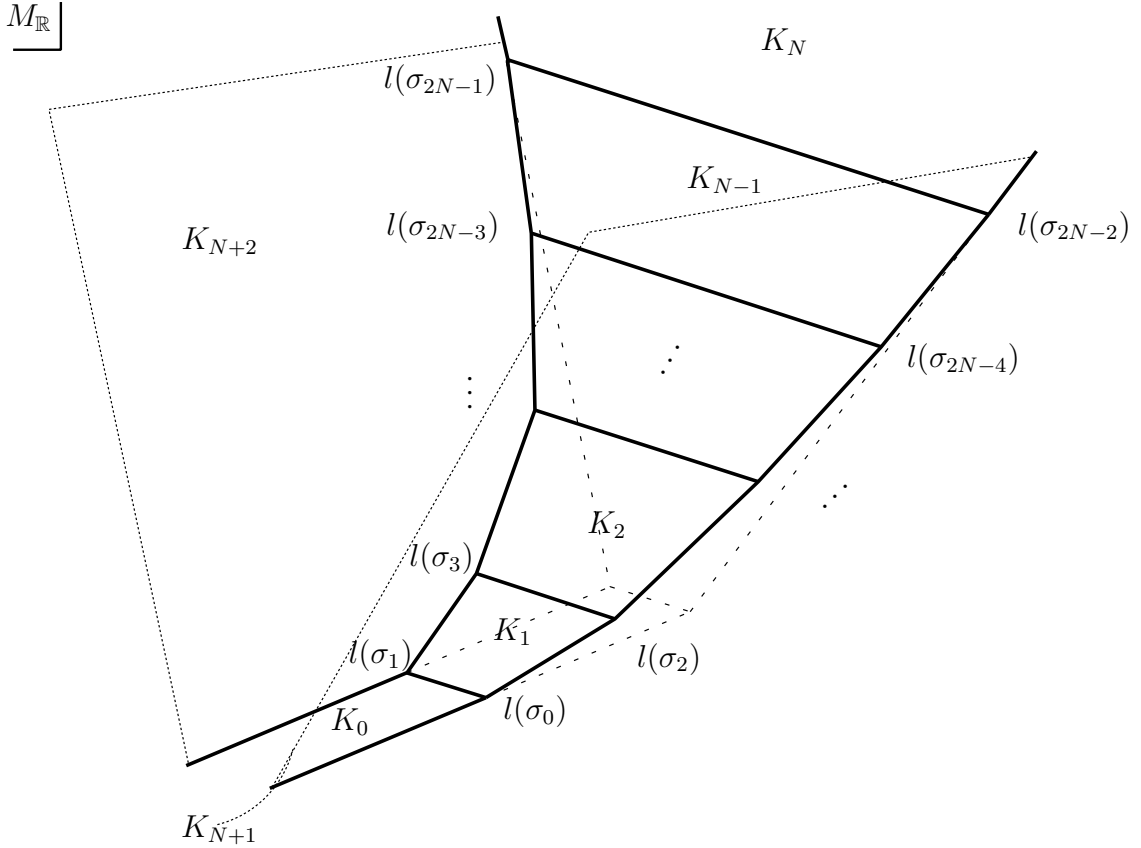}
  \caption{\textit{The polyhedron $\mathcal{P}(D)$}}
  \label{fig;PD_of_SU(N)}
 \end{center}
\end{figure}

Reflecting the noncompactness of the $SU(N)$ geometry, 
the polyhedron is unbounded. 
See Figure \ref{fig;PD_of_SU(N)}. 
There appear $N+3$ facets or faces of codimension one. 
These are the intersections with hyperplanes 
$K_i 
\equiv
\Bigl\{ x \in M_{\mathbb{R}}\,;\, 
\langle\,x,\, v_i\,\rangle+d_i=0 
\Bigr\}$.     
The Cartier divisor for the one-dimensional cones 
can be written as $l(\rho_i)=K_i \cap M$. 
The Cartier divisor for the three-dimensional cones is 
$l(\sigma_{2k})=K_k \cap K_{k+1} \cap K_{N+1}$ 
and 
$l(\sigma_{2k+1})=K_k \cap K_{k+1} \cap K_{N+2}$.   
These become the faces of codimension three.  
$\mathcal{P}(D)$ is a polyhedron with 
apex at each $l(\sigma_i)$. 
The Cartier divisor for the two-dimensional cones 
is $l(\gamma)=(K_j \cap K_k)\cap M$, where 
$\rho_j$ and $\rho_k$ are facets of the two-dimensional 
cone $\gamma$.  
The faces of codimension two are the intersections with 
these $K_j \cap K_k$.

\subsubsection{Counting the dimensions}

\begin{figure}[htb]
 \psfrag{0}{}
 \psfrag{1}{}
 \psfrag{2}{}
 \psfrag{3}{}
 \psfrag{4}{}
 \psfrag{5}{}
 \psfrag{6}{}
 \psfrag{7}{}
 \psfrag{8}{}
 \psfrag{9}{}
 \psfrag{k0}{$K_0$}
 \psfrag{k1}{}
 \psfrag{k2}{}
 \psfrag{k3}{}
 \psfrag{k4}{}
 \psfrag{k5}{$K_{N}$}
 \psfrag{k6}{$K_{N+1}$}
 \psfrag{k7}{$K_{N+2}$}
 \psfrag{o1}{$o_1$}
 \psfrag{o2}{$o_2$}
 \psfrag{M}{\hspace{-2mm}$M_{\mathbb{R}}$}
 \begin{center}
  \includegraphics[scale=.8]{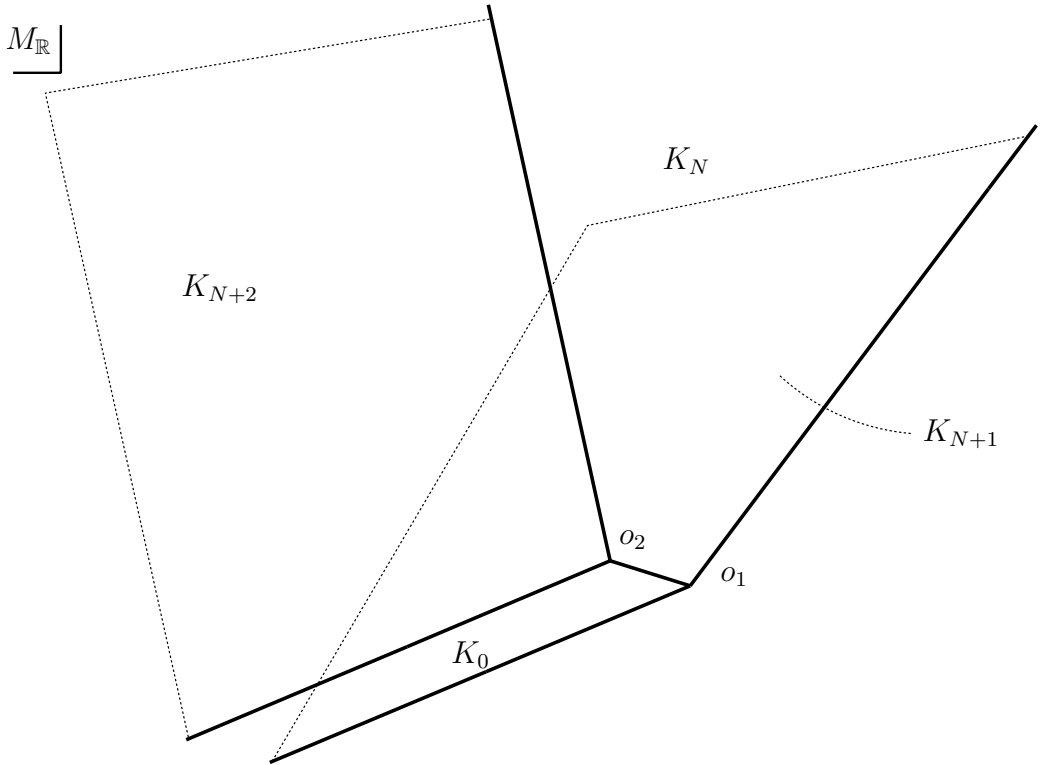}
  \caption{\textit{The polyhedron $\Theta (D)$}}
  \label{fig;Theta_D}
 \end{center}
\end{figure}

The physical Hilbert space $\mathcal{H}_{\omega}$ 
is infinite dimensional since the number of lattice points 
that are in $\mathcal{P}(D)$ is infinite. 
Instead, we will count the number of the lattice points 
in a relative manner. This may give us a finite value. 
We define a convex polyhedron in $M_{\mathbb{R}}$ by 
\begin{eqnarray}
\Theta(D)\equiv 
\Bigl\{ 
x \in M_{\mathbb{R}}\,;\, 
\langle \,x,\, v_i\, \rangle+d_i 
\geq 0 
\hspace{3mm} 
\mbox{for} 
\hspace{2mm}
\rho_{i}= 
\rho_0\,,\rho_{N}\,,\rho_{N+1}\,,\rho_{N+2} 
\Bigr\}\,.
\label{theta(D)}
\end{eqnarray}
The facets are the intersections with 
$K_0, K_N,K_{N+1}$ and $K_{N+2}$. 
See Figure \ref{fig;Theta_D}.
In particular, 
$\Theta(D)$ is a polyhedron with apexes at 
$o_1$ and $o_2$, where 
\begin{eqnarray}
o_1 
&\equiv& K_0 \cap K_N \cap K_{N+1}  
\nonumber \\
&=& 
(d_0-d_{N+1})e_1^*+\frac{d_0-d_N}{N}e_2^*-d_0e_3^*\,,
\nonumber \\
o_2 
&\equiv& 
K_0 \cap K_N \cap K_{N+2} \nonumber \\ 
&=& 
o_1+T_Ne_1^*\,. 
\label{o1 and o2}
\end{eqnarray}
Note that the polyhedron (\ref{theta(D)}) naturally appears 
if one considers line bundles on 
an unresolved $A_{N-1}$ singularity fibred over $\mathbb{P}^1$. 

\begin{figure}[htb]
 \psfrag{0}{$l(\sigma_0)$}
 \psfrag{1}{\hspace{-1mm}$l(\sigma_1)$}
 \psfrag{2}{$l(\sigma_2)$}
 \psfrag{3}{\hspace{-1mm}$l(\sigma_3)$}
 \psfrag{4}{\hspace{4mm}\rotatebox{45}{$\cdots$}}
 \psfrag{5}{\hspace{0mm}\rotatebox{90}{$\cdots$}}
 \psfrag{6}{$l(\sigma_{2N-4})$}
 \psfrag{7}{\hspace{-13mm}$l(\sigma_{2N-3})$}
 \psfrag{8}{$l(\sigma_{2N-2})$}
 \psfrag{9}{\hspace{-7mm}$l(\sigma_{2N-1})$}
 \psfrag{k0}{$K_0$}
 \psfrag{k1}{$K_1$}
 \psfrag{k2}{$K_2$}
 \psfrag{k3}{\rotatebox{60}{$\cdots$}}
 \psfrag{k4}{$K_{N-1}$}
 \psfrag{k5}{$K_{N}$}
 \psfrag{k6}{$K_{N+1}$}
 \psfrag{k7}{$K_{N+2}$}
 \psfrag{o1}{$o_1$}
 \psfrag{o2}{$o_2$}
 \psfrag{M}{\hspace{-2mm}$M_{\mathbb{R}}$}
\begin{center}
  \includegraphics[scale=.8]{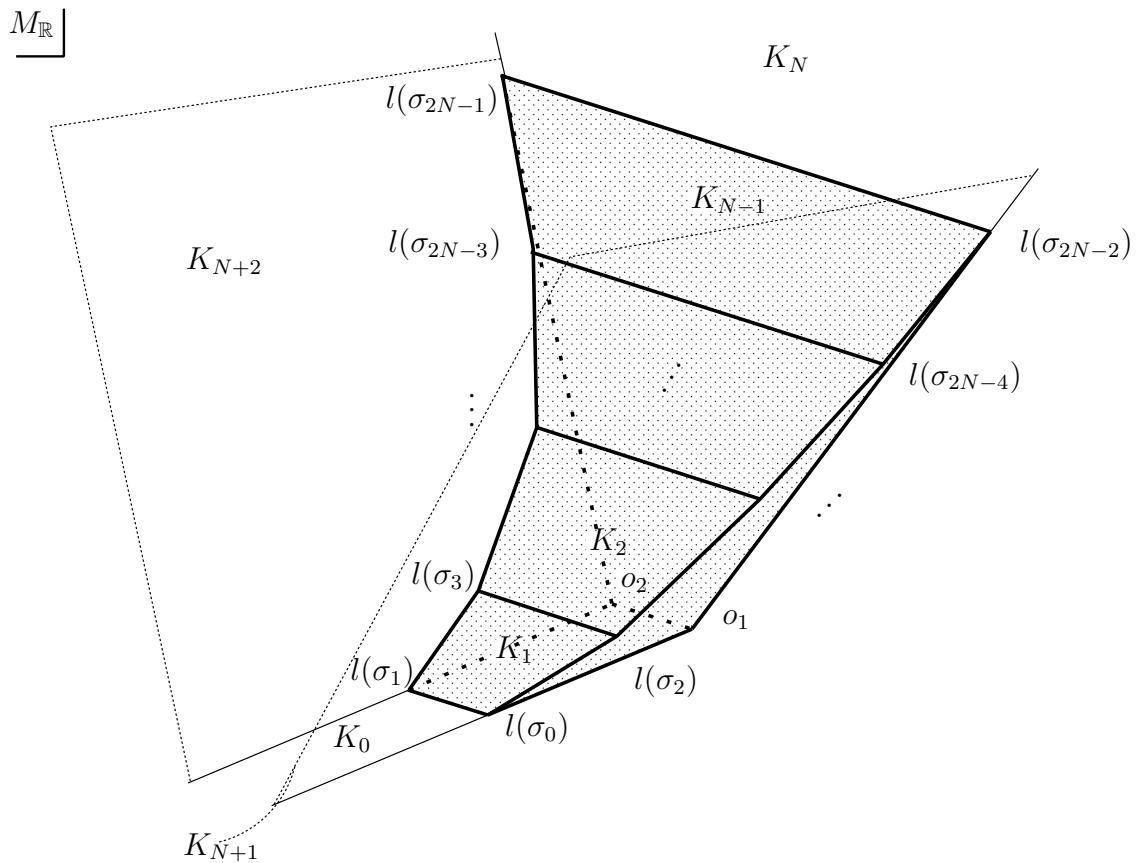}
  \caption{\textit{The complement $\mathcal{P}^c(D)$}}
 \label{PDc_of_SU(N)}
 \end{center}
\end{figure}

It is clear from the definitions 
that $\mathcal{P}(D) \subset \Theta(D)$. 
Let 
\begin{eqnarray}
\mathcal{P}^c(D) 
\equiv 
\Theta(D)\setminus \mathcal{P}(D)\,.  
\label{Pc(D)}
\end{eqnarray}
$\mathcal{P}^c(D)$ is bounded 
and the number of lattice points in it becomes finite. 
See Figure \ref{PDc_of_SU(N)}.
We introduce (regularized) dimensions of the physical 
Hilbert space $\mathcal{H}_{\omega}$ by the relative 
cardinality of $\mathcal{P}(D) \cap M$. 
\begin{eqnarray}
\mbox{dim}
\hspace{1mm}
\mathcal{H}_{\omega}
\equiv 
-\mbox{Card}
\bigl( \mathcal{P}^c(D) \cap M \bigr)\,,  
\label{definition of dim H}
\end{eqnarray}
where the minus sign is a convention. 

The lattice $M$ determines a volume element 
on $M_{\mathbb{R}}$, by letting the volume of 
the unit cube determined by a basis of $M$ be one. 
The volume of $\mathcal{P}^c(D)$ is related to 
the number of lattice points by 
\begin{eqnarray}
\mbox{Vol}(\mathcal{P}^c(D))
=
\lim_{\nu \rightarrow \infty}
\frac{\mbox{Card}\bigl(\mathcal{P}^c(\nu D) \cap M \bigr)}{\nu^3}\,.  
\label{Vol and Card}
\end{eqnarray}
Note that any scaling of $D$ can be 
translated to simultaneous scalings 
of $T_i$ by (\ref{Ti by di}) and is expressed as 
a scaling of $g_{st}$ by (\ref{quantized kahler parameters}). 
In particular, the limit $\nu \rightarrow \infty$ 
in (\ref{Vol and Card}) is converted to 
the semi-classical limit 
which is obtained by letting $g_{st} \rightarrow 0$.

\begin{figure}[htb]
 \psfrag{1}{$l(\sigma_{2k-2})$}
 \psfrag{2}{\hspace{-2mm}$l(\sigma_{2k-1})$}
 \psfrag{3}{$l(\sigma_{2k})$}
 \psfrag{4}{\hspace{-4mm}$l(\sigma_{2k+1})$}
 \psfrag{o1}{$o_{1}$}
 \psfrag{o2}{$o_{2}$}
 \begin{center}
  \includegraphics[scale=.8]{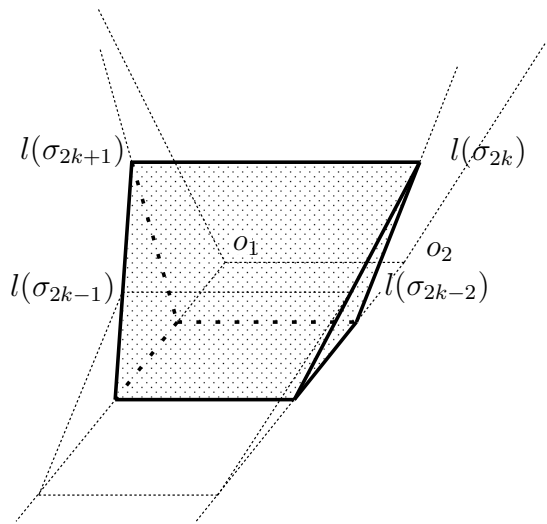}
  \caption{\textit{The pentahedron $\Delta_k$. 
           The complement $\mathcal{P}^c(D)$ is the sum of 
           $\Delta_k$.}}
  \label{fig;Delta_k}
 \end{center}
\end{figure}

In order to describe the regularized dimensions, 
taking the above into account, 
we compute the volume of $\mathcal{P}^c(D)$ 
rather than the number of lattice points in 
$\mathcal{P}^c(D)$. 
\begin{eqnarray}
\mbox{dim}
\hspace{1mm}
\mathcal{H}_{\omega}
=
-\mbox{Vol}(\mathcal{P}^c(D))
+\mbox{an error term}\,.  
\label{dim H by volume}
\end{eqnarray}
The error term is bounded by the two-dimensional area 
of the boundary of $\mathcal{P}^c(D)$ and vanishes 
at the semi-classical limit. 
We first divide $\mathcal{P}^c(D)$ into 
the pieces of pentahedra. 
The pentahedron $\triangle_k$ is surrounded by 
$K_0,K_k,K_{k+1},K_{N+1}$ and $K_{N+2}$. 
See Figure \ref{fig;Delta_k}. 
The volume is given by the sum 
$\sum_{k=1}^{N-1} \mbox{Vol}(\triangle_k)$. 
The volume of each pentahedron can be computed as follows. 
\begin{eqnarray}
\mbox{Vol}(\triangle_k)=
\frac{1}{3k(k+1)}
\Bigl( \sum_{i=1}^kiT_i \Bigr)^3 
+
\frac{T_N}{2k(k+1)} 
\Bigl( \sum_{i=1}^k iT_i \Bigr)^2\,. 
\label{volume of triangle k}
\end{eqnarray}
Therefore we obtain 
\begin{eqnarray}
\mbox{Vol} (\mathcal{P}^c(D))
=
\sum_{k=1}^{N-1}
\frac{1}{3k(k+1)}
\Bigl( \sum_{i=1}^kiT_i \Bigr)^3 
+T_N
\sum_{k=1}^{N-1}\frac{1}{2k(k+1)} 
\Bigl( \sum_{i=1}^k iT_i \Bigr)^2\,. 
\label{volume of Pc(D)}
\end{eqnarray}

\subsubsection{The dimensions vs. gauge theory prepotential}

The regularized dimensions (\ref{definition of dim H}) 
turns to be a perturbative prepotential for five-dimensional 
supersymmetric gauge theory at the semi-classical limit.

It becomes convenient to express 
the quantized K$\ddot{\mbox{a}}$hler parameters 
for the fibred ALE space by using $U(1)$ charges as follows. 
\begin{eqnarray}
T_k = p_{N-k+1}-p_{N-k} 
\hspace{1cm} 
(k=1,2,\,\cdots\,,N-1)\,, 
\label{Ti by U(1) charges} 
\end{eqnarray}
where $\bigl\{ p_r \bigr\}_{r=1}^N$ are 
$U(1)$ charges which are ordered as 
$p_1 \leq p_{2} \leq \cdots \leq p_N$. 
We will associate these charges with partitions 
in the next section. 
We further impose the $SU(N)$ condition on 
the charges. 
\begin{eqnarray}
\sum_{r=1}^Np_r=0\,. 
\label{SU(N) condition}
\end{eqnarray}
This condition slightly restricts 
allowed values of the quantized K$\ddot{\mbox{a}}$hler 
parameters. In particular, 
it requires $\sum_{i=1}^{N-1}iT_i \in N \mathbb{Z}$. 
By using (\ref{Ti by di}), 
this is rephrased as $d_0-d_N \in N \mathbb{Z}$. 
Hence, by (\ref{o1 and o2}), $o_1 \in M$.

The geometric engineering \cite{geometric engineering} 
dictates that four-dimensional $\mathcal{N}=2$ supersymmetric 
$SU(N)$ Yang-Mills is realized by the $SU(N)$ geometry. 
In particular, 
the classical K$\ddot{\mbox{a}}$hler parameter 
$t_B \equiv g_{st}T_N$ 
of the base $\mathbb{P}^1$ is proportional to 
$1/g^2$, where $g$ is the gauge coupling constant at the 
string scale. The gauge instanton effect is weighted 
with $e^{-c_2/g^2}\sim \Lambda^{2Nc_2}$, where $\Lambda$ 
is the scale parameter of the gauge theory. 
This leads to $e^{-t_B}\sim \Lambda^{2N}$. 
The classical K$\ddot{\mbox{a}}$hler parameters $t_i=g_{st}T_i$ 
of the blow-up cycles in the fibre are proportional to $a_r$, 
which are the VEVs of the adjoint scalar in the vector multiplet.

A possible relation between the quantized $SU(N)$ geometry 
and five-dimensional $\mathcal{N}=1$ supersymmetric $SU(N)$ 
Yang-Mills is suggested in \cite{MNTT} through the study of 
a statistical model of random plane partitions. 
The geometry parameters are converted to 
the gauge theory parameters by the following 
identification \cite{MNTT}. 
\begin{eqnarray}
g_{st}=2R\hbar\,, 
\hspace{6mm} 
\widetilde{p_r}=a_r/\hbar \,,  
\hspace{6mm} 
g_{st}T_N=-2N \ln (2R\Lambda)\,,  
\label{geometry and gauge theory parameters}
\end{eqnarray}
where $\widetilde{p}_r\equiv p_r+\xi_r$ with 
$\xi_r= \frac{1}{N}(r-\frac{N+1}{2})$. 
The parameter $\hbar$ plays the similar role as $g_{st}$ 
in the statistical model, and $R$ is the radius of $S^1$ 
in the fifth dimension. $a_r$ and $\Lambda$ are as above.

Let us put $a_{rs}\equiv a_r-a_s$. 
The identification 
(\ref{geometry and gauge theory parameters}) gives 
\begin{eqnarray}
\sum_{i=1}^{k}iT_i=
\frac{1}{\hbar}\sum_{r=N-k+1}^Na_{r\,N-k}
-\frac{k(k+1)}{2N}\,. 
\end{eqnarray} 
By using this, 
the following $\hbar$-expansions are obtainable 
after some computations. 
\begin{eqnarray}
\sum_{k=1}^{N-1}
\frac{1}{2k(k+1)}
\Bigl( \sum_{i=1}^kiT_i \Bigr)^2 
&=& 
\frac{1}{\hbar^2}
\Bigl\{ 
\frac{1}{2N}\sum_{r>s}^Na_{rs}^2+O(\hbar) 
\Bigr\}\,, 
\nonumber \\
\sum_{k=1}^{N-1}\frac{1}{3k(k+1)} 
\Bigl( \sum_{i=1}^k iT_i \Bigr)^3 
&=& 
\frac{1}{\hbar^3} 
\Bigl\{ 
\frac{1}{6}\sum_{r>s}^Na_{rs}^3
-\frac{N}{6}\sum_{r=1}^Na_r^3 
+O(\hbar) 
\Bigr\}\,. 
\label{habr-expansions of T2 and T3}
\end{eqnarray} 
These lead to  
\begin{eqnarray}
g_{st} \cdot 
\mbox{Vol}(\mathcal{P}^c(D)) 
=\frac{1}{\hbar^2} 
\Bigl\{ 
\mathcal{F}_{\mbox{{\scriptsize  5dSYM}}}^{pert}(a,\,\Lambda,\,R)
+O(\hbar) 
\Bigr\}\,, 
\label{volume as prepotential}
\end{eqnarray}
where  
\begin{eqnarray}
\mathcal{F}^{pert}_{\mbox{{\scriptsize 5dSYM}}}
(a,\,\Lambda,\,R)=
2R
\left(
\frac{1}{6}\sum_{r>s}^Na_{rs}^3
-\frac{N}{6}\sum_{r=1}^Na_r^3 
\right)
-\ln (2R\Lambda)
\sum_{r>s}^Na_{rs}^2  
\label{perturbative prepotential of SU(N) theory with k=N}
\end{eqnarray}
is the perturbative prepotential 
for five-dimensional $\mathcal{N}=1$ supersymmetric 
$SU(N)$ Yang-Mills with the Chern-Simons term 
\cite{Seiberg,triple intersection}. 
The five-dimensional theory is living on 
$\mathbb{R}^4 \times S^1$ at the large radius limit, 
and the Chern-Simons coupling constant 
$k_{\mbox{\tiny{C.S.}}}$ is quantized to N.

Therefore, by (\ref{dim H by volume}) 
and (\ref{volume as prepotential}), we finally obtain 
\begin{eqnarray}
g_{st} \cdot 
\mbox{dim}\, \mathcal{H}_{\omega}
=
-\frac{1}{\hbar^2} 
\Bigl\{ 
\mathcal{F}^{pert}_{\mbox{{\scriptsize 5dSYM}}}
(a,\,\Lambda,\,R)
+O(\hbar) 
\Bigr\}\,. 
\label{dims as prepotential}
\end{eqnarray}

\subsection{Geometric quantization and quantum foam}
\label{geometric quantization and quantum foam}

The geometric quantization converts the $SU(N)$ geometry 
into foam of gravitational quanta. 
Let us explain this by using the gauged linear 
$\sigma$-model description rather than algebraic toric 
geometry.

Consider $\mathbb{C}^{N+3}$, 
equipped with the standard symplectic form 
$\omega_{\mathbb{C}^{N+3}}=
\sqrt{-1}\sum_{i=0}^{N+2}d\bar{z}_i \wedge dz_i$, 
where $z=(z_0,\, \cdots\,,z_{N+2})$ denote the coordinates. 
Consider the following 
$U(1)^{\otimes N}$ action on $\mathbb{C}^{N+3}$. 
\begin{eqnarray}
z_i \longmapsto  
z_i\prod_{a=1}^N\exp(\sqrt{-1}Q_{ia}\theta_a)\,,  
\label{NU(1)action}
\end{eqnarray}
where $Q=(Q_{ia})$ is the charge matrix 
(\ref{Q}) with $\widetilde{k}_{\mbox{\tiny{C.S.}}}=0$. 
The action is a hamiltonian action generated by the momentum 
map $\Phi=(\phi_1,\, \cdots\,, \phi_N)$ given by  
\begin{eqnarray}
\phi_a(z)
\equiv \sum_{i=0}^{N+2}
Q_{ia}|z_i|^2\,. 
\label{momentum map Phi}
\end{eqnarray}
Let $X_t$ be the K$\ddot{\mbox{a}}$hler quotient taken at 
$t=(t_1,\,\cdots\,,t_{N})$.  
\begin{eqnarray}
X_t \equiv 
\frac{\,\,\,\Phi^{-1}(t)\,\,\,}{U(1)^{\otimes N}}\,. 
\label{Xt}
\end{eqnarray}
It becomes a Calabi-Yau threefold. 
When all $t_a$ are positive, 
it is the $SU(N)$ geometry with 
the classical K$\ddot{\mbox{a}}$hler parameters $t$.

$X_t$ is a (singular) fibration of three-dimensional real torus. 
To see this, 
we start with the standard torus action 
on $\mathbb{C}^{N+3}$. $(N+3)$-dimensional 
real torus acts on $\mathbb{C}^{N+3}$ by   
$z_i \mapsto e^{\sqrt{-1}\theta_i}z_i$. 
This is again a hamiltonian action generated by the momentum map 
$u=(u_0,\,\cdots \,,u_{N+2})$ given by  
\begin{eqnarray}
u_i(z)\equiv |z_i|^2\,. 
\label{momentum map u}
\end{eqnarray}
The image of $\mathbb{C}^{N+3}$ by the momentum map becomes 
$\bigl( \mathbb{R}_{\geq 0} \bigr)^{\oplus N+3}$. 
A commutative diagram 
\begin{eqnarray}
\begin{array}{ccc}
\mathbb{C}^{N+3}
&
\begin{array}{c}
\mbox{\small{$\Phi$}} 
\\[-4mm]
-\hspace{-2mm}-\hspace{-2mm}-\hspace{-2mm}
\longrightarrow 
\\[-4mm]
~
\end{array}
&
\mathbb{R}^N 
\\[-3mm]
u \downarrow 
& 
~
& 
\parallel 
\\[-2mm]
\bigl( \mathbb{R}_{\geq 0} \bigr)^{\oplus N+3}
&
\begin{array}{c}
~
\\[-4mm]
-\hspace{-2mm}-\hspace{-2mm}-\hspace{-2mm}
\longrightarrow 
\\[-3mm]
\mbox{\small{$\Phi_{\geq}$}}
\end{array}
&
\mathbb{R}^N 
\end{array}
\label{commutative diagram}
\end{eqnarray}
defines $\Phi_{\geq}$ as 
\begin{eqnarray}
\Phi_{\geq}(u) \equiv 
\sum_{i=0}^{N+2}Q_{ia}u_i\,. 
\label{real Phi}
\end{eqnarray}
Let $P_t$ be the inverse image of $t$. 
\begin{eqnarray}
P_t &\equiv& \Phi_{\geq}^{-1}(t) 
\nonumber \\
&=& 
\Bigl\{ 
(u_0,\,\cdots\,,u_{N+2}) 
\in 
\bigl( \mathbb{R}_{\geq 0} \bigr)^{\oplus N+3}\,;
\hspace{2mm}
\sum_{i=0}^{N+2}Q_{ia}u_i=t_a\, 
\Bigr\}\,.
\label{Pt}
\end{eqnarray}
It is a three-dimensional real manifold with corners. 
The commutativity of the diagram (\ref{commutative diagram}) 
implies the following fibration 
of $(N+3)$-dimensional real torus. 
\begin{eqnarray}
\Phi^{-1}(t)\,\,
\stackrel{\mbox{\small{$u$}}}\longrightarrow\,\,
P_t\,. 
\label{(N+3)d torus fibration}
\end{eqnarray}
It is clear that each fibre is stable by the $U(1)^{\otimes N}$ 
action (\ref{NU(1)action}). By taking the quotient fibrewise 
in (\ref{(N+3)d torus fibration}) we obtain 
\begin{eqnarray}
X_t 
\,\,
\stackrel{\mbox{\small{$\pi$}}}\longrightarrow\,\,
\,\,
P_t\,. 
\label{3d torus fibration}
\end{eqnarray}
This describes the torus fibration of $X_t$.   
Singular fibres (lower dimensional tori) are on the 
boundary of $P_t$. 
Three-dimensional real torus $T_{\mathbb{R}}$ 
acts on each fibre, 
where $T_{\mathbb{R}}$ is the compact part of the algebraic 
torus $T$. This torus action becomes a hamiltonian action 
generated by the projection $\pi$ of the fibration.  
It is the momentum map of the torus action.

We can make a connection with the previous algebraic 
toric description.  
Let $\omega$ be a quantized K$\ddot{\mbox{a}}$hler form on the 
$SU(N)$ geometry $X_{\Delta}$.   
The quantized K$\ddot{\mbox{a}}$hler parameters $T$ 
are set to be positive integrals.  
The classical K$\ddot{\mbox{a}}$hler parameters are given by 
$t=g_{st}T$. 
The K$\ddot{\mbox{a}}$hler quotient $X_{t}$ becomes identical to 
$(X_{\Delta}, \omega)$. 
The geometric quantization can be understood as follows. 
Recall that {\it minimal volume} is the least volume 
measurable quantum mechanically in a phase space.
The minimal volume becomes $\sim g_{st}^3$ in the present quantization.  
We may divide $X_{t}$ into pieces of the minimal volume. 
Each piece is interpreted as a gravitational quantum. 
Thus $X_t$ is discretized into a foam of gravitational quanta. 
The number of quanta is estimated as  
$\mbox{Vol}(X_t)/g_{st}^3
=\frac{1}{3!}\int_{X_{\Delta}}(\omega/g_{st})^3$. 
The geometric quantization shows that 
these quanta are actually labeled by $\mathcal{P}(D) \cap M$.

\begin{figure}[htb]
 \psfrag{1}{\hspace{-2mm}$m_0$}
 \psfrag{2}{$m_1$}
 \psfrag{3}{$m_2$}
 \psfrag{4}{$m_3$}
 \psfrag{5}{\rotatebox{60}{$\cdots$}}
 \psfrag{6}{\rotatebox{120}{$\cdots$}}
 \psfrag{7}{\hspace{-2mm}$m_{2N-4}$}
 \psfrag{8}{$m_{2N-3}$}
 \psfrag{9}{\hspace{-2mm}$m_{2N-2}$}
 \psfrag{10}{$m_{2N-1}$}
 \psfrag{o1}{$\widehat{o_{1}}$}
 \psfrag{o2}{$\widehat{o_{2}}$}
 \psfrag{o}{0}
 \psfrag{ex}{$e_x$}
 \psfrag{ey}{$e_y$}
 \psfrag{ez}{$e_z$}
 \psfrag{L}{$L_{\mathbb{R}}$}
 \begin{center}
  \includegraphics[scale=.8]{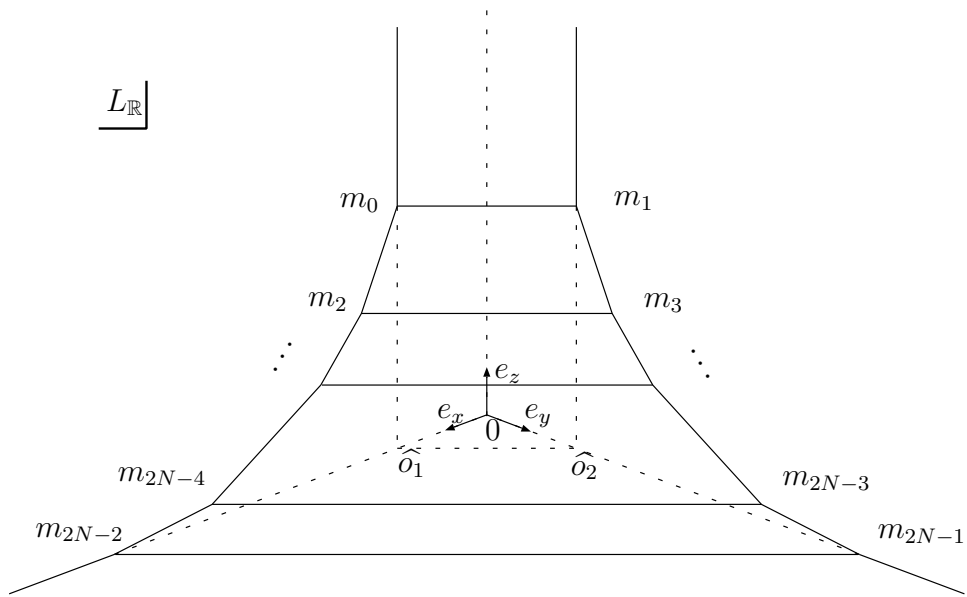}
  \caption{\textit{The polyhedron
  $P_T$  in 
  $L_{\mathbb{R}_{\geq 0}}$}}
  \label{fig;PT}
 \end{center}
\end{figure}

By rescaling $u_i$ to $u_i/g_{st}$, 
the real manifold $P_t$ will be identified 
with the polyhedron $\mathcal{P}(D)$, 
where $c_1(\mathcal{O}(D))=[\omega]/g_{st}$. 
The minimal volume is naturally brought about from  
the uncertainty relation 
$\Delta q_i \cdot \Delta p_i \sim g_{st}$, where 
$z_i=q_i+\sqrt{-1}p_i$. 
The uncertainty relation 
also leads to a discretization of $P_t$ by making  
$u_i/g_{st} \in \mathbb{Z}_{\geq 0}$. 
Thus, it is naturally expected that 
the discretization is identical to 
$\mathcal{P}(D) \cap M$. 
Let us confirm this.  
We consider the real manifold $P_T$ 
rather than $P_t$ after the rescalings. 
We also project $P_T$ into the three-dimensional space 
$(u_N,u_{N+1},u_{N+2})$ by solving the conditions 
$\sum_{i}Q_{ia}u_i=T_a$ in terms of these variables. 
Let $L$ denote a lattice generated by $e_x,e_y$ and $e_z$. 
\begin{eqnarray}
L=\mathbb{Z}e_x+\mathbb{Z}e_y+\mathbb{Z}e_z\,. 
\label{lattice L}
\end{eqnarray}
We identify the three-dimensional space with $L_{\mathbb{R}}$ by 
\begin{eqnarray}
(u_N,u_{N+1},u_{N+2})\, 
\longmapsto \,
u_{N+1}e_x+u_{N+2}e_y+u_{N}e_z\,. 
\end{eqnarray}
It turns out that $P_T$ is realized by a polyhedron in 
$L_{\mathbb{R}_{\geq 0}}\equiv \mathbb{R}_{\geq 0}e_x+
\mathbb{R}_{\geq 0}e_y+\mathbb{R}_{\geq 0}e_z$, 
as depicted in Figure \ref{fig;PT}.  
It has apexes at $m_0,m_1,\cdots,m_{2N-1} \in L$, where 
\begin{eqnarray}
m_{2k}
&=& 
\bigl(
T_N+2\sum_{i=1}^kiT_i 
\bigr)e_x
+
\sum_{i=k+1}^{N}(N-i)T_ie_z\,, 
\nonumber \\
m_{2k+1}
&=& 
\bigl(
T_N+2\sum_{i=1}^kiT_i 
\bigr)e_y
+
\sum_{i=k+1}^{N}(N-i)T_ie_z\,. 
\label{m_k}
\end{eqnarray}

The discretization of $P_T$ 
is achieved in $L_{\mathbb{R}}$ by 
a sublattice of $L$ that is obtained by the projection.   
This sublattice turns to be identified 
with the dual lattice $M$. 
The identification is conveniently described by 
using a linear transformation. 
Let $\iota$ be a linear transformation from 
$M_{\mathbb{R}}$ to $L_{\mathbb{R}}$ defined by 
\begin{eqnarray}
\iota(u)=A(u-l(\sigma_0))+m_0\, 
\hspace{7mm} u \in M_{\mathbb{R}}\,, 
\label{iota}
\end{eqnarray}
where $A$ is  given by 
\begin{eqnarray}
A(e_1^*)=-e_x+e_y, 
\hspace{5mm}
A(e_2^*)=Ne_z, 
\hspace{5mm}
A(e_3^*)=e_x+e_y+e_z.  
\label{matrix A}
\end{eqnarray}
Then the lattice is given by $\iota(M)$. 
It is a degree $2N$ sublattice of $L$.  
We also see that $P_T$ is the image of 
$\mathcal{P}(D)$. 
\begin{eqnarray}
P_T=\iota(\mathcal{P}(D))\,. 
\label{P(D) vs P_T}
\end{eqnarray}
In particular we have 
$m_i=\iota(l(\sigma_i))$ 
for each $i$. 
Therefore the discretization satisfies 
\begin{eqnarray}
P_T \cap \iota(M)= \iota (\mathcal{P}(D) \cap M)\,.  
\label{discrete P_T}
\end{eqnarray}

The polyhedron $\Theta(D)$ is mapped to $P_{(0,\cdots,0,T_N)}$ 
by $\iota$. It becomes a polyhedron 
in $L_{\mathbb{R}_{\geq 0}}$,  
with apexes at $\hat{o}_1=T_Ne_x$ and 
$\hat{o}_2=T_Ne_y$. See Figure \ref{fig;PT}. 
The corresponding threefold $X_{(0,\cdots,0,t_N)}$ is the singular 
Calabi-Yau threefold with $A_{N-1}$ singularity along the fibres. 
Therefore the complement $\mathcal{P}^c(D)$ satisfies  
\begin{eqnarray}
\iota(\mathcal{P}^c(D))
=P_{(0,\cdots,0,T_N)}\setminus P_{T}\,, 
\label{P_c(D) vs P_0/P_T}
\end{eqnarray}
and 
\begin{eqnarray}
\iota(\mathcal{P}^c(D) \cap M)=
\,
\bigl( 
P_{(0,\cdots,0,T_N)}\setminus P_{T}
\bigr) 
\,
\cap 
\iota(M)\,. 
\label{quantized P_c(D) vs P_0/P_T}
\end{eqnarray}

\subsection{Gauge$\slash$gravity correspondence 
for arbitrary framing}
\label{gauge/gravity correspondence 2}

We now generalize the discussion in section 
\ref{gauge/gravity correspondence 1} to the case 
of local $SU(N)$ geometry with arbitrary framing $k_{\mbox{\tiny{C.S.}}}$.
Let $\omega$ be a quantized K$\ddot{\mbox{a}}$hler form 
on local $SU(N)$ geometry. 
Let $D=\sum_{i=0}^{N+2}d_iD_i$ be the corresponding divisor. 
The quantized K$\ddot{\mbox{a}}$hler parameters are expressed 
by using (\ref{quantized kahler parameters}) 
and (\ref{c1 numbers of O(D)}) as  
\begin{eqnarray}
 T_k 
  &=& 
  d_{k-1}-2d_k+d_{k+1} 
  \hspace{8mm}
(1 \leq k \leq N-1)\,, 
\nonumber \\
 T_N
  &=& 
  (\widetilde{k}_{\mbox{\tiny{C.S.}}}-2)d_0
-\widetilde{k}_{\mbox{\tiny{C.S.}}}d_1 +d_{N+1}+d_{N+2}
  \, .
\label{Ti by di_k}
\end{eqnarray}

The physical Hilbert space $\mathcal{H}_{\omega}$ 
is given by the global sections of $\mathcal{O}(D)$. 
These are labeled by $\mathcal{P}(D) \cap M$, where  
$\mathcal{P}(D)$ is a rational convex polyhedron 
defined by the same equations as (\ref{PD}).  
Since $\mathcal{H}_{\omega}$ is infinite dimensional, 
we will count the dimensions in a relative manner  
as is done for $\widetilde{k}_{\mbox{\tiny{C.S.}}}=0$. 
We notice that, when $\widetilde{k}_{\mbox{\tiny{C.S.}}}\neq 0$, 
the polyhedron $\Theta(D)$ becomes a polyhedron 
with apexes at $o_1$ and $o_2$ such that  
\begin{eqnarray}
o_1 
&=& K_0 \cap K_N \cap K_{N+1}  
\nonumber \\
&=& 
(d_0-d_{N+1})e_1^*+\frac{d_0-d_N}{N}e_2^*-d_0e_3^*\,,
\nonumber \\
o_2 
&=& 
K_0 \cap K_N \cap K_{N+2} \nonumber \\ 
&=& 
o_1+T_B e_1^*\,,  
\label{o1 and o2_k}
\end{eqnarray}
where $T_B$ has been introduced by 
\begin{eqnarray}
T_B=T_N-\frac{\widetilde{k}_{\mbox{\tiny{C.S.}}}}{N}
  \sum\limits_{i=1}^{N-1} (N-i)T_i  
\label{TB}
\end{eqnarray}
and required to be nonnegative. 
We define the regularized dimensions similarly 
by (\ref{definition of dim H}). 
At the semi-classical limit it becomes  
\begin{eqnarray}
\mbox{dim}
\hspace{1mm}
\mathcal{H}_{\omega}= 
-\mbox{Vol}(\mathcal{P}^c(D))+\mbox{an error term}\,. 
\label{dim H for k}
\end{eqnarray}
The volume can be computed as follows.
\begin{eqnarray}
 \mbox{Vol} (\mathcal{P}^c(D))
  &=&
  \sum_{k=1}^{N-1}
  \frac{1}{3k(k+1)}
  \Bigl( \sum_{i=1}^kiT_i \Bigr)^3 
  +T_B
  \sum_{k=1}^{N-1}\frac{1}{2k(k+1)} 
  \Bigl( \sum_{i=1}^k iT_i \Bigr)^2 \nonumber \\
 &&
  +\frac{\widetilde{k}_{\mbox{\tiny{C.S.}}}}{6}
  \sum\limits_{k=1}^{N}
  \Bigl\{
   \frac{1}{N}\sum\limits_{i=1}^{N-1}
   \left(N-i\right)T_i
   -\sum\limits_{i=1}^{k-1}T_i
  \Bigr\}^3 \, .
\label{volume of Pc(D)_k}
\end{eqnarray}
The effect of the framing appears explicitly in the third term and 
implicitly through $T_B$ in the second term.

We can also present a gauge theory description of the dimensions. 
We first impose the $SU(N)$ condition 
(\ref{SU(N) condition}) on the quantized K$\ddot{\mbox{a}}$hler 
parameters for the fibre. This makes $T_B \in \mathbb{Z}_{\geq 0}$. 
We will translate the geometry parameters into the gauge theory 
parameters by generalizing the identification 
(\ref{geometry and gauge theory parameters}) as follows. 
\begin{eqnarray}
g_{st}=2R\hbar\,, 
\hspace{6mm} 
\widetilde{p_r}=a_r/\hbar \,,  &
\hspace{6mm} 
g_{st}T_B=-2N \ln (2R\Lambda)\,. 
\label{geometry and gauge theory parameters_k}
\end{eqnarray}
The identification leads to 
(\ref{habr-expansions of T2 and T3}) and 
\begin{eqnarray}
 \sum\limits_{k=1}^{N}
  \Bigl\{
   \frac{1}{N}\sum\limits_{i=1}^{N-1}
   \left(N-i\right)T_i
   -\sum\limits_{i=1}^{k-1}T_i
  \Bigr\}^3
  &=& 
  \frac{1}{\hbar^3} 
  \Bigl\{ 
  \sum\limits_{r=1}^{N}
  a_r^3
  +O(\hbar) 
  \Bigr\}\,.
\label{habr-expansions of T2 and T3_k}
\end{eqnarray} 
It follows from these that 
the $\hbar$-expansion of the volume 
(\ref{volume of Pc(D)_k}) becomes 
\begin{eqnarray}
g_{st} \cdot 
\mbox{Vol}(\mathcal{P}^c(D)) 
=\frac{1}{\hbar^2} 
\Bigl\{ 
\mathcal{F}_{\mbox{{\scriptsize  5dSYM}}}^{pert}(a,\,\Lambda,\,R)
+O(\hbar) 
\Bigr\}\,, 
\label{volume as prepotential_k}
\end{eqnarray}
where  
\begin{eqnarray}
\mathcal{F}^{pert}_{\mbox{{\scriptsize 5dSYM}}}
(a,\,\Lambda,\,R)=
2R
\left(
\frac{1}{6}\sum_{r>s}^Na_{rs}^3
-\frac{k_{\mbox{\tiny{C.S.}}}}{6}\sum_{r=1}^Na_r^3 
\right)
-\ln (2R\Lambda)
\sum_{r>s}^Na_{rs}^2  
\label{perturbative prepotential with k}
\end{eqnarray}
is the perturbative prepotential for five-dimensional 
$\mathcal{N}=1$ supersymmetric $SU(N)$ Yang-Mills plus  
the Chern-Simons term with the coupling constant $k_{\mbox{\tiny{C.S.}}}$.

\subsection{Perspective from the Riemann-Roch-Hirzebruch theorem}
\label{Riemann-Roch}

We argue the validity of the regularized dimensions 
by showing how the definition does fit to the Riemann-Roch formula.

Let $X$ be a compact nonsingular $n$-dimensional 
algebraic variety. 
Let $\mathcal{O}(D)$ be the line bundle of a divisor $D$. 
The Euler-Poincar$\acute{\mbox{e}}$ characteristic of the 
line bundle is defined by 
\begin{eqnarray}
\chi(X,\mathcal{O}(D))= 
\sum_{p}\,(-)^p\, \mbox{dim}\, H^p(X,\mathcal{O}(D))\,. 
\label{Euler-Poincare}
\end{eqnarray}
The Riemann-Roch-Hirzebruch theorem says 
\begin{eqnarray}
\chi(X,\mathcal{O}(D))=
\int_X \mbox{Td}(X) \mbox{ch}(\mathcal{O}(D))\,, 
\label{R-R formula}
\end{eqnarray}
where $\mbox{ch}(\mathcal{O}(D))$ denotes the Chern 
character of the line bundle and $\mbox{Td}(X)$ 
is the Todd genus of X. 
These characteristic classes have a homological 
representation by the Poincar$\acute{\mbox{e}}$ duality.  
For instance, $c_1(\mathcal{O}(D))$ is dual to $D$. 
By using the homological representation, one can deduce 
from (\ref{R-R formula}):   
\begin{eqnarray}
\lim_{\nu \rightarrow \infty}
\frac{\chi(X,\mathcal{O}(\nu D))}{\nu^n}
=\frac{D^n}{n!}\,, 
\label{Euler by Dn}
\end{eqnarray}
where $D^n$ denotes the self-intersection number 
obtained by intersecting $D$ with itself $n$ times. 
When all the higher cohomology groups vanish, 
the above becomes 
\begin{eqnarray}
\lim_{\nu \rightarrow \infty}
\frac{1}{\nu^n}\,\mbox{dim}\,H^0(X,\mathcal{O}(\nu D))
=
\frac{D^n}{n!}\,.
\label{H0 by Dn}
\end{eqnarray}

Although local $SU(N)$ geometry is a noncompact 
Calabi-Yau threefold, we would like to establish 
a formula analogous to (\ref{H0 by Dn}) 
on the dimensions (\ref{definition of dim H}). 
Let $\omega$ be a quantized K$\ddot{\mbox{a}}$hler form 
on local $SU(N)$ geometry. Let $\mathcal{O}(D)$ be 
the corresponding line bundle of 
$c_1(\mathcal{O}(D))=[\omega]/g_{st}$. 
We impose the $SU(N)$ condition (\ref{SU(N) condition}) 
on the quantized K$\ddot{\mbox{a}}$hler class. 
When this condition is satisfied, 
the following divisor can be chosen 
among the linear equivalence class. 
\begin{eqnarray}
D=\sum_{i=1}^{N-1}d_iD_i+d_{N+2}D_{N+2}\,. 
\label{SU(N) D}
\end{eqnarray}

Let us compute the triple self-intersection number 
of the above divisor. 
Triple intersection numbers involving the compact divisors 
$D_i$ ($1 \leq i \leq N-1$) besides the noncompact divisor  
$D_{N+2}$ can be calculated by the standard manner. 
Those which do not vanish are listed as follows. 
\begin{eqnarray}
\begin{array}{ccc}
D_i^3=8, 
\hspace{3mm} &   
D_i^2 \cdot D_{i+1}= -2(i+1)+\widetilde{k}_{\mbox{\tiny{C.S.}}}, & 
\hspace{3mm} 
D_i \cdot D_{i+1}^2= 2i-\widetilde{k}_{\mbox{\tiny{C.S.}}}, \\[1mm]
D_i^2 \cdot D_{N+2}= -2, 
\hspace{3mm} &
D_i \cdot D_{i+1}\cdot D_{N+2}=1 ,
\hspace{3mm} &
D_{N+2}^3=x . \\
\end{array}  
\label{DiDjDk}
\end{eqnarray}
All the other combinations vanish. In particular, 
$D_i \cdot D_{N+2}^2=0$. The self-intersection number 
$D_{N+2}^3$ is not unique and set to be $x$. 
By using (\ref{DiDjDk}), the triple self-intersection 
number becomes 
\begin{eqnarray}
\frac{D^3}{3!}
&=& 
\frac{4}{3}
\sum_{i=1}^{N-1}d_i^3
+\sum_{i=1}^{N-2}
 \Bigl(\frac{\widetilde{k}_{\mbox{\tiny{C.S.}}}}{2}-i-1\Bigr)
      d_{i+1}d_i^2
+\sum_{i=1}^{N-2}
 \Bigl(i-\frac{\widetilde{k}_{\mbox{\tiny{C.S.}}}}{2}\Bigr)
      d_{i+1}^2d_i 
  \nonumber \\
&& 
-\Bigl(
\sum_{i=1}^{N-1}d_i^2-\sum_{i=1}^{N-2}d_{i+1}d_i 
 \Bigr)d_{N+2}
+\frac{x^3}{6}d_{N+2}^3\,. 
\label{triple intersection by di}
\end{eqnarray}

It is possible to rewrite  
(\ref{triple intersection by di}) in terms of 
the quantized K$\ddot{\mbox{a}}$hler parameters. 
Note that we are putting 
$d_0=d_N=d_{N+1}=0$ by linear equivalence. 
It follows from (\ref{Ti by di_k}) that the quantized 
K$\ddot{\mbox{a}}$hler parameters satisfy 
\begin{eqnarray} 
\sum_{i=1}^kiT_i
=-(k+1)d_k+kd_{k+1}\,. 
\end{eqnarray}
By using this, we can see 
\begin{eqnarray}
\sum_{k=1}^{N-1}
\frac{1}{2k(k+1)} 
\Bigl( \sum_{i=1}^kiT_i \Bigr)^2 
&=& 
\sum_{i=1}^{N-1}d_i^2
-\sum_{i=1}^{N-2}d_{i+1}d_i\,, 
\nonumber   \\
\sum_{k=1}^{N-1}
\frac{1}{3k(k+1)} 
\Bigl( \sum_{i=1}^k iT_i \Bigr)^3 
&=& 
-\frac{4}{3}
\sum_{i=1}^{N-1}d_i^3
+\sum_{i=1}^{N-2}(i+1)d_{i+1}d_i^2
-\sum_{i=1}^{N-2}id_{i+1}^2d_i 
\nonumber \, ,\\
\sum\limits_{i=1}^{N}
\Bigl\{
\frac{1}{N}\sum\limits_{n=1}^{N-1}
\left(N-n\right)T_n
-\sum\limits_{n=1}^{i-1}T_n
\Bigr\}^3 
&=&
3\sum\limits_{i=1}^{N-1}
\Bigl(
d_{i+1}d_i^2-d_{i+1}^2d_i
\Bigr)  \,.
\label{T2 and T3 by di}
\end{eqnarray}
By comparing (\ref{triple intersection by di}) 
with (\ref{volume of Pc(D)_k}) taking account of 
(\ref{T2 and T3 by di}), we find 
\begin{eqnarray}
\frac{D^3}{3!}=
-\mbox{Vol}
\bigl( \mathcal{P}^c(D) \bigr)
+\frac{x}{6}T_B^3\,. 
\label{voume of Pc(D) by D}
\end{eqnarray}

Therefore, 
it follows from (\ref{definition of dim H}) and  
(\ref{Vol and Card}) that the corresponding regularized 
dimensions satisfies 
\begin{eqnarray}
\lim_{\nu \rightarrow \infty}
\frac{1}{\nu^3}\,\mbox{dim}\,
\mathcal{H}_{\nu \omega} 
=
\frac{D^3}{3!}-\frac{x}{6}T_B^3\,. 
\label{dim H by D3}
\end{eqnarray}

\section{K$\ddot{\mbox{a}}$hler Gravity 
Seen by Random Plane Partitions}

The ground state of a certain model of random plane partitions 
leads to the quantum foam of the local geometry.
The ground state energy is essentially given 
by the dimensions of the physical Hilbert space of 
the K$\ddot{\mbox{a}}$hler gravity.  
Each gravitational quantum of the geometry 
consists of $N$ cubes of plane partitions.

\subsection{A model of random plane partitions}
\label{a model of random plane partition}

\begin{figure}[hbt]
 \begin{center}
  \includegraphics{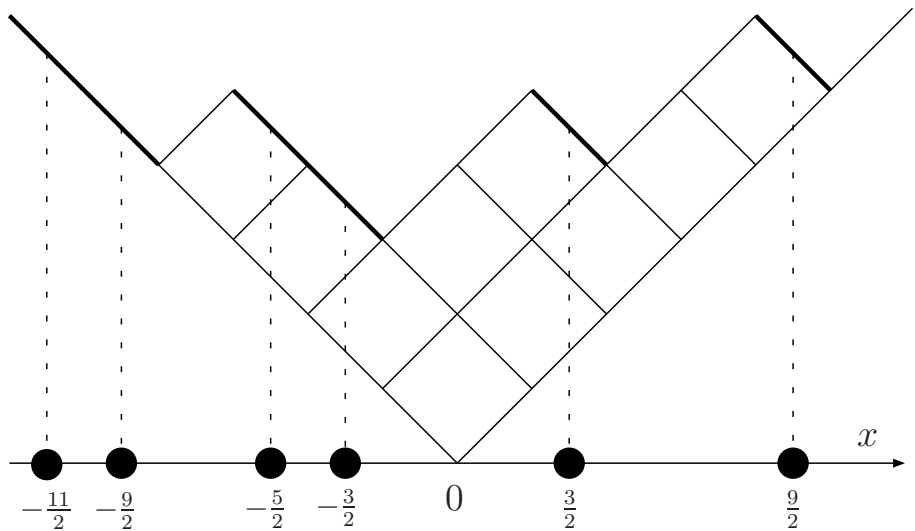}
 \end{center}
 \caption{\textit{The Young diagram and the Maya diagram 
          for $\mu=(5,3,1,1)$}} 
 \label{fig;maya}
\end{figure}

A partition $\mu$ is a sequence of non-negative integers
satisfying $\mu_{i} \geq \mu_{i+1}$ for all $i\geq 1$.
Partitions are often identified with the Young diagrams.
The size of $\mu$ is defined by $|\mu|=\sum_{i \geq 1}\mu_i$.  
It is the total number of boxes of the diagram.
It is also known that partitions 
are identified with the Maya diagrams. 
The Maya diagram $\mu$ is a sequence of strictly 
decreasing numbers $x_i(\mu)\equiv \mu_i-i+\frac{1}{2}$ for 
$i \geq 1$. The correspondence with the Young diagram 
is depicted in Figure \ref{fig;maya}. 

\begin{figure}[tb]
 \psfrag{pi0}{$\pi(0)$}
 \psfrag{m}{m}
 \psfrag{0}{0}
 \psfrag{pi}{$\pi_{ij}$}
 \psfrag{i}{$i$}
 \psfrag{j}{$j$}
 \begin{center}
 \includegraphics{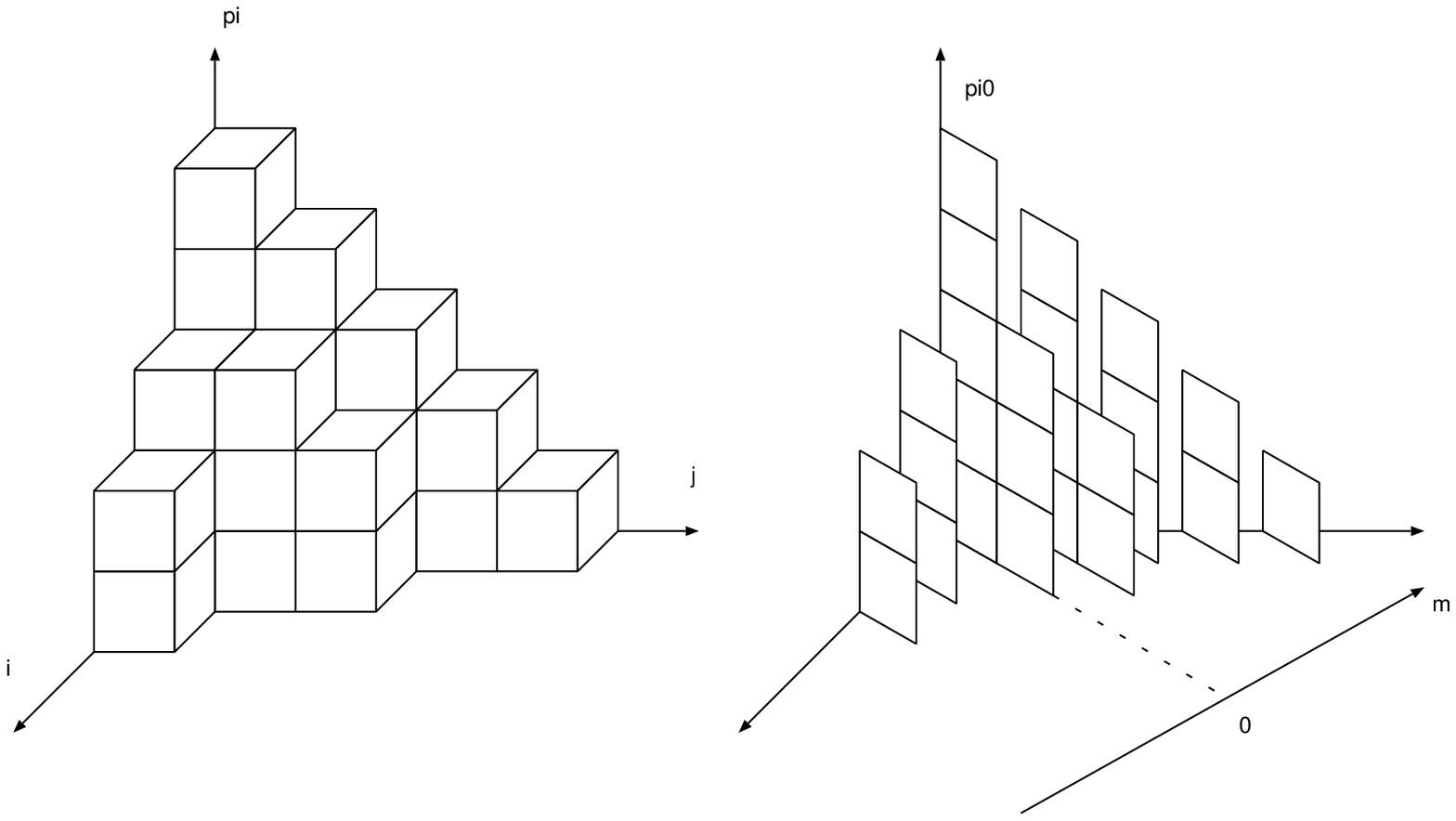}
 \caption{
\textit{The three-dimensional Young diagram and its diagonal slices}} 
 \label{fig;plane_part}
 \end{center}
\end{figure}

A plane partition $\pi$ is an array of 
non-negative integers satisfying 
$\pi_{ij}\geq \pi_{i+1 j}$ and $\pi_{ij}\geq \pi_{i j+1}$ 
for all $i,j \geq 1$. 
Plane partitions are identified 
with the three-dimensional 
Young diagrams. The three-dimensional diagram $\pi$ 
is a set of unit cubes such that $\pi_{ij}$ cubes 
are stacked vertically on each $(i,j)$-element of $\pi$. 
See Figure \ref{fig;plane_part}.
The size of $\pi$ is defined by 
$|\pi|=\sum_{i,j \geq 1}\pi_{ij}$,  
which is the total number of cubes of the diagram. 
Diagonal slices of $\pi$ become partitions.    
Let $\pi(m)$ denote the partition 
along the $m$-th diagonal slice. 
In particular $\pi(0)$ is the main diagonal partition. 
See also Figure \ref{fig;plane_part}. 
The series of partitions $\pi(m)$ satisfies the condition
\begin{eqnarray}
\cdots \prec \pi(-2) \prec \pi(-1) \prec 
\pi(0) \succ \pi(1) \succ \pi(2) \succ \cdots,
\label{time evolution}
\end{eqnarray}
where $\mu \succ \nu$ means the following interlace relation 
between two partitions $\mu$ and $\nu$.  
\begin{eqnarray}
\mu \succ \nu ~~~
\Longleftrightarrow ~~~
\mu_1 \geq \nu_1 \geq \mu_2 \geq \nu_2 
\geq \mu_3 \geq \cdots.
\end{eqnarray}

We first consider a statistical model of 
plane partitions defined by the following 
partition function.
\begin{eqnarray}
Z(q,Q)
&\equiv& 
\sum_{\pi}\, q^{|\pi|}\,Q^{|\pi(0)|}\,, 
\label{Z(q,Q)}
\end{eqnarray}
where $q$ and $Q$ are indeterminate.   
The Boltzmann weight consists of two parts. 
The first contribution comes from the energy 
of plane partitions, and 
the second contribution is a chemical potential 
for the main diagonal partitions.

The condition (\ref{time evolution}) suggests that 
plane partitions are certain evolutions of partitions 
by the discretized time $m$. 
This leads to a hamiltonian formulation 
for the statistical model.  
In particular, the transfer matrix approach 
developed in \cite{Okounkov-Reshetikhin} 
express the partition function (\ref{Z(q,Q)}) 
in terms of two-dimensional conformal field theory. 
Let us recall that 
partitions can be mapped to the Fock space of 
two-dimensional free fermions by 
\begin{eqnarray}
|\mu;p \rangle =
 \Bigl(
 \prod\limits_{i=1}^{l(\mu)} 
 \psi_{-x_i(\mu)-p}
 \Bigr)
 \Bigl(
 \prod\limits_{i=1}^{l(\mu)} 
 \psi^*_{-i+\frac{1}{2}+p}
 \Bigr) 
|\emptyset;p\rangle\,,     
\label{eq;c_part_state}
\end{eqnarray}
where $l(\mu)$ denotes the length of $\mu$, 
that is, the number of the non-zero $\mu_i$, 
and the state $|\emptyset ; p\rangle$ is 
the ground state for the charge $p$ sector. 
See appendix for details on the free fermion system. 
Combinations of partitions with the $U(1)$ charges are called 
charged partitions. Such a charged partition is denoted 
by $(\mu,p)$, where $\mu$ is a partition 
and $p$ is the $U(1)$ charge. 
To describe the transfer matrix, 
it is convenient to introduce vertex operators  
\begin{eqnarray}
\Gamma_{+}(z) 
&\equiv& 
\exp 
  \Bigl(   
      \sum\limits_{k=1}^{\infty}  
      \frac{1}{k}z^{k}J_k   
  \Bigr)\,, 
\nonumber \\
\Gamma_{-}(z) 
&\equiv& 
\exp 
  \Bigl(
     \sum\limits_{k=1}^{\infty}
     \frac{1}{k}z^{-k}J_{-k}
  \Bigr)\,   . 
\label{eq;transfer}
\end{eqnarray}
where $J_k$ are the modes of the $U(1)$ current.
These operators satisfy the following properties. 
\begin{eqnarray}
\langle \mu;p | \Gamma_{+}(z) |\lambda;p\rangle 
&=&
\left\{
 \begin{array}{cc}
   z^{-|\mu|+|\lambda|} & \mbox{$\mu\prec\lambda$} \\
   0 & \mbox{otherwise}\,.
 \end{array} 
\right. 
\nonumber \\
\langle \mu;p |\Gamma_{-}(z)|\lambda;p\rangle 
&=&
\left\{
 \begin{array}{cc}
    z^{-|\mu|+|\lambda|} & \mbox{$\mu\succ\lambda$} \\
    0 & \mbox{otherwise}\,.
 \end{array} 
\right. 
\label{eq;transfer2}
\end{eqnarray}
By comparing (\ref{eq;transfer2}) with (\ref{time evolution}), 
we see that these operators can provide a description 
of the evolutions of partitions. 
The evolution at a negative time $m$ is given  
by $\Gamma_+(q^{-(m+\frac{1}{2})})$,  
while the evolution at a nonnegative time $m$ 
is by $\Gamma_-(q^{-(m+\frac{1}{2})})$. 
The partition function has the following 
expression by using the vertex operators.
\begin{eqnarray}
Z(q,Q)=
\langle \emptyset;0 | 
\Bigl(
 \prod\limits_{m=-\infty}^{-1}
 \Gamma_{+}(q^{-(m+\frac{1}{2})})
\Bigr)\,
Q^{L_0}\,
\Bigl(
 \prod\limits_{m=0}^{\infty} 
 \Gamma_{-}(q^{-(m+\frac{1}{2})})
\Bigr)
|\emptyset;0\rangle\,. 
\label{eq;top_u(1)}
\end{eqnarray}

We can interpret the random plane partitions  
as a $q$-deformation of random partitions. 
It may be seen by rewriting (\ref{Z(q,Q)}) as 
\begin{eqnarray}
Z(q,Q)=\sum_{\mu}\,Q^{|\mu|}\, 
\Bigl( \sum_{\pi(0)=\mu}\, q^{|\pi|} \Bigr)\,. 
\label{Z(q,Q) random partitions}
\end{eqnarray}
The partitions $\mu$ are thought 
as the ensemble of the model by summing first over  
plane partitions whose main diagonal partitions are $\mu$. 
In the transfer matrix description 
this is obtained by factorizing 
the amplitude (\ref{eq;top_u(1)}) at $m=0$. 
The factorization by using the charged partition states 
(\ref{eq;c_part_state}) gives rise to 
\begin{eqnarray}
Z(q,Q)=
\sum_{\mu}Q^{|\mu|}
s_{\mu}(q^{\frac{1}{2}},q^{\frac{3}{2}},\cdots)^2, 
\label{Z U(1) schur}
\end{eqnarray}
where $s_{\mu}(q^{\frac{1}{2}},q^{\frac{3}{2}},\cdots)$ 
is the Schur function 
$s_{\mu}(x_1,x_2,\cdots)$ specialized at 
$x_i=q^{i-\frac{1}{2}}$ \cite{Macdonald}.

\subsection{Random plane partitions 
and supersymmetric gauge theories}

The Fock representation of a single complex fermion 
has \cite{Miwa-Jimbo} an alternative realization 
that is obtained by exploiting $N$ component complex fermions 
$\psi^{(r)}(z)$, $\psi^{*(s)}(z)$ 
according to the identification.  
\begin{eqnarray}
\psi_{k}^{(r)} = \psi_{N(k-\xi_{r})}\,, 
\hspace{8mm}
\psi_{l}^{(s)*} = \psi_{N(l+\xi_{s})}\,,
\label{N component fermion}
\end{eqnarray}
where $k,l\in\mathbb{Z}+\frac{1}{2}$ and $r,s=1,2,\ldots,N$.  
This allows us to express a charged partition $(\mu,p)$ 
uniquely by means of $N$ charged partitions $(\lambda^{(r)},p_r)$ 
and vice versa, 
through the realizations of the charged partition state. 
\begin{eqnarray}
|\mu; p\rangle =
\bigotimes_{r=1}^{N} 
|\lambda^{(r)}; p_r\rangle_{(r)}\,. 
\label{N-1 states}
\end{eqnarray}
The equality can be read 
in terms of the characteristic functions 
(\ref{characteristic function for charged partition}) 
for charged partitions as follows.  
\begin{eqnarray}
P_{\mu,\,p}(t)=\sum_{r=1}^Ne^{N\xi_rt}P_{\lambda^{(r)},\,p_r}(Nt)\,.   
\label{N-1 character}
\end{eqnarray} 
By applying the method of power-sums \cite{B-O} 
one can deduce the following information from 
(\ref{N-1 character}).
\begin{eqnarray}
p &=& 
\sum\limits_{r=1}^{N} p_r\,, 
\label{N-1 charges} \\
|\mu| &=& 
N \sum\limits_{r=1}^{N} |\lambda^{(r)}|
+\frac{N}{2}\sum\limits_{r=1}^{N} p_r^2
+\sum\limits_{r=1}^{N}rp_r\,, 
\label{N-1 sizes} \\
\kappa(\mu) &=& 
N^2\sum\limits_{r=1}^{N}
\kappa(\lambda^{(r)})
+2N^2\sum\limits_{r=1}^{N}
    \widetilde{p}_r|\lambda^{(r)}|
+\frac{N^2}{3}\sum\limits_{r=1}^{N}
\widetilde{p}_r^3\,, 
\label{N-1 torsions}
\end{eqnarray}
where $\kappa(\mu)$ measures asymmetry of a partition 
and is defined by 
$2\kappa(\mu)\equiv 
\sum\limits_{i \geq 1}\mu_i^2-\sum\limits_{i \geq 1}\,^{t}\mu^{2}_i$.  
Here $^{t}\mu$ denotes the partition conjugate to $\mu$, 
which is obtained by flipping the Young diagram $\mu$ 
over its main diagonal.

The factorization (\ref{Z(q,Q) random partitions}) 
or (\ref{Z U(1) schur}) can be also expressed 
in terms of $N$ charged partitions, 
by using (\ref{N-1 states}).
Let us factor the partition function into 
\begin{eqnarray}
Z(q,Q)= 
\sum_{p_r}
Z_{SU(N)}(p_r;q,Q)\,, 
\label{factored random plane partitions}
\end{eqnarray}
where the $U(1)$ charges $p_r$ automatically satisfy 
the $SU(N)$ condition (\ref{SU(N) condition}), 
due to the charge conservation (\ref{N-1 charges}). 
We have included the summation over partitions 
$\lambda^{(r)}$ implicitly in $Z_{SU(N)}(p_r;q,Q)$.

The factorization 
(\ref{factored random plane partitions}) 
turns to be a bridge between the random plane partitions 
and supersymmetric gauge theories. 
We identify the parameters $q,Q$ and $p_r$ 
with the gauge theory parameters as follows. 
\begin{eqnarray}
q=e^{-\frac{2}{N}R\hbar},
\hspace{4mm}
Q=(2R\Lambda)^{2}, 
\hspace{4mm}
\widetilde{p}_r=a_r/\hbar\,. 
\label{mapping RPP to gauge theory}
\end{eqnarray}
It is shown in \cite{MNTT} that the above identification 
leads to 
\begin{eqnarray}
Z_{SU(N)}(p_r;q,Q)=
Z_{5d\, \mbox{\scriptsize{SYM}}}(a_r;\Lambda,R,\hbar)\,, 
\label{exact partition function for SU(N)SYM}
\end{eqnarray}
where the RHS is the exact partition function \cite{N-O} 
for five-dimensional $\mathcal{N}=1$ supersymmetric $SU(N)$ 
Yang-Mills plus the Chern-Simons term 
having the coupling constant equal to $N$.

Since the K$\ddot{\mbox{a}}$hler gravity on 
local $SU(N)$ geometry has the relation 
to the perturbative dynamics of 
the gauge theory, we take a closer look of 
the perturbative contribution 
in (\ref{exact partition function for SU(N)SYM}). 
A charged ground partition is a charged partition    
determined by a set of $N$ charged empty partitions 
$(\emptyset, p_r)$ by using (\ref{N-1 states}). 
The neutral ground partition is simply 
called ground partition and is denoted by 
$\mu_{\mbox{\tiny{GP}}}(p_r)$. 
A set $\mathcal{M}(p_r)$ is a set of plane partitions 
such that 
\begin{eqnarray}
\mathcal{M}(p_r)\equiv 
\Bigl\{ 
\pi\,\, ; \, \pi(0)=\mu_{\mbox{\tiny{GP}}}(p_r) 
\Bigr\}\,. 
\label{M(pi_GP)}
\end{eqnarray}
The perturbative contribution can be described \cite{MNTT} 
by random plane partitions restricted within $\mathcal{M}(p_r)$.  
\begin{eqnarray}
Z_{SU(N)}^{pert}(p_r;q,Q)=
Q^{\left|\mu_{\mbox{\tiny{GP}}}(p_r)\right|}
\sum_{\,\,\pi \in \mathcal{M}(p_r)}
q^{|\pi|}\,. 
\label{perturbative partition function for SU(N)SYM}
\end{eqnarray}
This turns out to be written by using the Schur function 
as follows. 
\begin{eqnarray}
Z_{SU(N)}^{pert}(p_r;q,Q)=
Q^{\left|\mu_{\mbox{\tiny{GP}}}(p_r)\right|}
s_{\mu_{\mbox{\tiny{GP}}}(p_r)}
(q^{\frac{1}{2}},\,q^{\frac{3}{2}},\,\cdots)^2\,. 
\end{eqnarray}

\subsection{Ground plane partition and 
quantum foam of local $SU(N)$ geometry}

\begin{figure}[hbt]
  \psfrag{a11}{$\pi_{11}$}
  \psfrag{a12}{$\pi_{12}$}
  \psfrag{a13}{$\pi_{13}$}
  \psfrag{a14}{$\pi_{14}$}
  \psfrag{a15}{$\pi_{15}$}
  \psfrag{a21}{$\pi_{21}$}
  \psfrag{a22}{$\pi_{22}$}
  \psfrag{a23}{$\pi_{23}$}
  \psfrag{a24}{$\pi_{24}$}
  \psfrag{a25}{$\pi_{25}$}
  \psfrag{a31}{$\pi_{31}$}
  \psfrag{a32}{$\pi_{32}$}
  \psfrag{a33}{$\pi_{33}$}
  \psfrag{a34}{$\pi_{34}$}
  \psfrag{a35}{$\pi_{35}$}
  \psfrag{a41}{$\pi_{41}$}
  \psfrag{a42}{$\pi_{42}$}
  \psfrag{a43}{$\pi_{43}$}
  \psfrag{a44}{$\pi_{44}$}
  \psfrag{a45}{$\pi_{45}$}
  \psfrag{a51}{$\pi_{51}$}
  \psfrag{a52}{$\pi_{52}$}
  \psfrag{a53}{$\pi_{53}$}
  \psfrag{a54}{$\pi_{54}$}
  \psfrag{a55}{$\pi_{55}$}
  \begin{center}
  \includegraphics{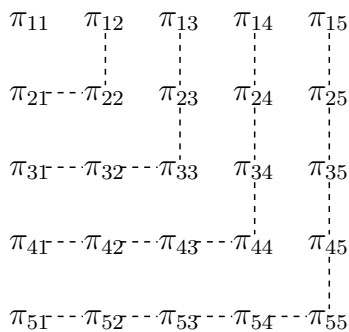}
  \caption{\textit{The minimizations determine the ground plane partition 
           by equating $\pi_{ij}$ that are connected by the dashed lines.}} 
  \label{fig;GPP}
  \end{center}
 \end{figure}

We consider the ground state of the statistical model 
(\ref{perturbative partition function for SU(N)SYM}). 
Such a ground state is called ground plane partition and is denoted by 
$\pi_{\mbox{\tiny{GPP}}}(p_r)$. 
It is a plane partition in $\mathcal{M}(p_r)$ 
which satisfies  
\begin{eqnarray}
|\pi_{\mbox{\tiny{GPP}}}(p_r)|\, \leq \, |\mu| 
\hspace{6mm}
\mbox{for}
\hspace{3mm}
\forall \mu \in \mathcal{M}(p_r)\,. 
\label{least action condition}
\end{eqnarray}
The above condition uniquely determines 
the ground plane partition. 
We regard plane partitions as the series of partitions. 
A series of partitions $\pi(m)$ gives an element of 
$\mathcal{M}(p_r)$ only when it evolves from/to 
$\pi(0)=\mu_{\mbox{\tiny{GP}}}(p_r)$ satisfying 
the interlace condition (\ref{time evolution}) 
at each discrete time. 
Since the Boltzmann weight is written as 
$|\pi|=\sum\limits_{m}|\pi(m)|$, 
the least action can be realized by the series 
that minimizes $|\pi(m)|$ at each $m$ 
from among partitions allowed 
by the interlace condition. 
The minimizations starting at $m=\pm 1$ determine 
the series recursively. See Figure \ref{fig;GPP}. 
These give rise to 
\begin{eqnarray}
\pi_{\mbox{\tiny{GPP}}}(p_r)(m)_i=
\mu_{\mbox{\tiny{GP}}}(p_r)_{i+|m|}\,. 
\label{GPP}
\end{eqnarray}
We note that the ground plane partition is actually 
the ground state of the model 
(\ref{exact partition function for SU(N)SYM}) 
and is regarded as a classical trajectory of 
the hamiltonian dynamics. 

\subsubsection{Perturbative gauge theory 
prepotential from ground plane partition}

\begin{figure}[hbt]
 \psfrag{v1}{\hspace{-3mm}$(N-1)T_1$}
 \psfrag{v2}{\hspace{-3mm}$(N-2)T_2$}
 \psfrag{v3}{$\vdots$}
 \psfrag{vN}{\hspace{-5mm}$T_{N-1}$}
 \psfrag{h1}{$T_1$}
 \psfrag{h2}{$2T_2$}
 \psfrag{h3}{$\cdots$}
 \psfrag{hN}{$(N-1)T_{N-1}$}
 \begin{center}
 \includegraphics{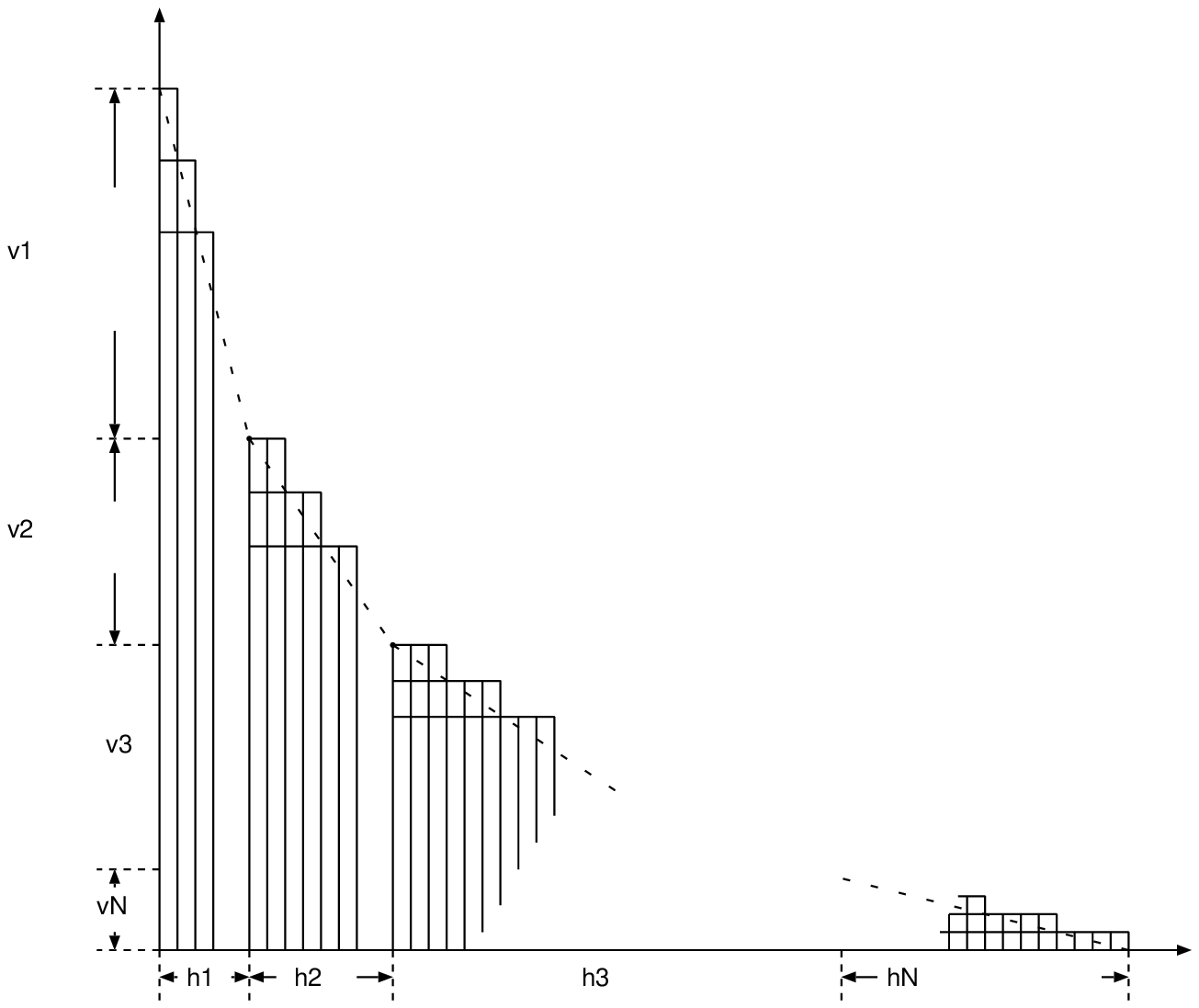}
 \caption{\textit{The Young diagram 
          $\mu_{\mbox{\tiny{GP}}}(p_r)$}} 
\label{fig;GP}
\end{center}
\end{figure}

We order the $U(1)$ charges such that 
$p_1 \leq p_2 \leq \cdots \leq p_N$. 
We first describe the ground partition explicitly.  
Use of $T_k$ becomes convenient rather than $p_r$ 
in the description. 
These were the quantized K$\ddot{\mbox{a}}$hler parameters 
in the previous section and are related 
with the $U(1)$ charges by (\ref{Ti by U(1) charges}). 
The ground partition becomes a partition of length 
$\sum\limits_{r=1}^{N-1}rT_r$ and is given by  
\begin{eqnarray}
\mu_{\mbox{\tiny{GP}}}
(p_r)_{r(m-1)+n+\sum\limits_{k=1}^{r-1}kT_k}=
-(N-r)(m-1)+\sum\limits_{k=r}^{N-1}(N-k)T_k\,, 
\label{GP}
\end{eqnarray}
where 
$1 \leq r \leq N-1,\,1 \leq m \leq T_r$ and 
$1 \leq n \leq r$. The corresponding Young diagram is 
depicted in Figure \ref{fig;GP}. 
Note that the size of ground partition can be read from 
(\ref{N-1 sizes}) and is expressed by means of $T_k$ as follows. 
\begin{eqnarray}
|\mu_{\mbox{\tiny{GP}}}(p_r)|
&=& 
N\sum_{k=1}^{N-1}\frac{1}{2k(k+1)}
\Bigl(\sum_{i=1}^kiT_i\Bigr)^2
+
\sum_{k=1}^{N-1}\frac{k(N-k)}{2}T_k\,. 
\label{size of GP}
\end{eqnarray}

Let us compute the size of ground plane partition. 
It follows from (\ref{GPP}) that the size is represented as 
\begin{eqnarray}
|\pi_{\mbox{\tiny{GPP}}}(p_r)|
&=& 
\sum_{i \geq 1}~
^t\mu_{\mbox{\tiny{GP}}}(p_r)_i^2\,. 
\label{size of GPP by GP}
\end{eqnarray} 
The RHS can be computed by using (\ref{GP}) and we obtain 
\begin{eqnarray}
|\pi_{\mbox{\tiny{GPP}}}(p_r)|=
N\sum_{k=1}^{N-1}\frac{1}{3k(k+1)}
\Bigl(\sum_{i=1}^kiT_i\Bigr)^3
+\sum_{k=1}^{N-1}\frac{1}{2}
\Bigl(\sum_{i=1}^kiT_i\Bigr)^2
+\sum_{k=1}^{N-1}\frac{k^2(N-k)}{6}T_k\,. 
\label{size of GPP}
\end{eqnarray}

We consider the Boltzmann weight for ground plane partition. 
We first note that, 
by combining the identifications 
(\ref{geometry and gauge theory parameters}) and 
(\ref{mapping RPP to gauge theory}), 
the parameters $q$ and $Q$ in the statistical model 
have the following expression.  
\begin{eqnarray}
q=e^{-g_{st}/N}, 
\hspace{5mm}
Q=q^{T_N}\,. 
\label{(q,Q) by (g_st,T_N)}
\end{eqnarray}
This allows us to write the Boltzmann weight as 
\begin{eqnarray}
q^{|\pi|}\,Q^{|\pi(0)|}=
q^{|\pi|+T_N|\pi(0)|}\,. 
\label{BW by (q,T_N)}
\end{eqnarray}
The exponent for ground plane partition is obtained from 
(\ref{size of GP}) and (\ref{size of GPP}). By comparing 
with (\ref{volume of Pc(D)}) we find  
\begin{eqnarray}
|\pi_{\mbox{\tiny{GPP}}}(p_r)|
+T_N|\mu_{\mbox{\tiny{GP}}}(p_r)|
=N \times \mbox{Vol}(\mathcal{P}^c(D))+O(T^2)\,. 
\label{Boltzmann wt vs volume}
\end{eqnarray}
Therefore the ground state energy gives rise to 
the gauge theory prepotential 
(\ref{perturbative prepotential of SU(N) theory with k=N}).
\begin{eqnarray}
-\ln\, 
q^{|\pi_{\mbox{\tiny{GPP}}}(p_r)|}\,
Q^{|\mu_{\mbox{\tiny{GP}}}(p_r)|}
=
\frac{1}{\hbar^2} 
\Bigl\{ 
\mathcal{F}^{pert}_{\mbox{{\scriptsize 5dSYM}}}
(a,\,\Lambda,\,R)
+O(\hbar) 
\Bigr\}\,. 
\end{eqnarray}

\begin{figure}[p]
  \begin{center}
   \begin{tabular}{ll}
    \begin{minipage}{90mm}
     \psfrag{(a)}{$(a)$}
     \includegraphics[scale=.6]{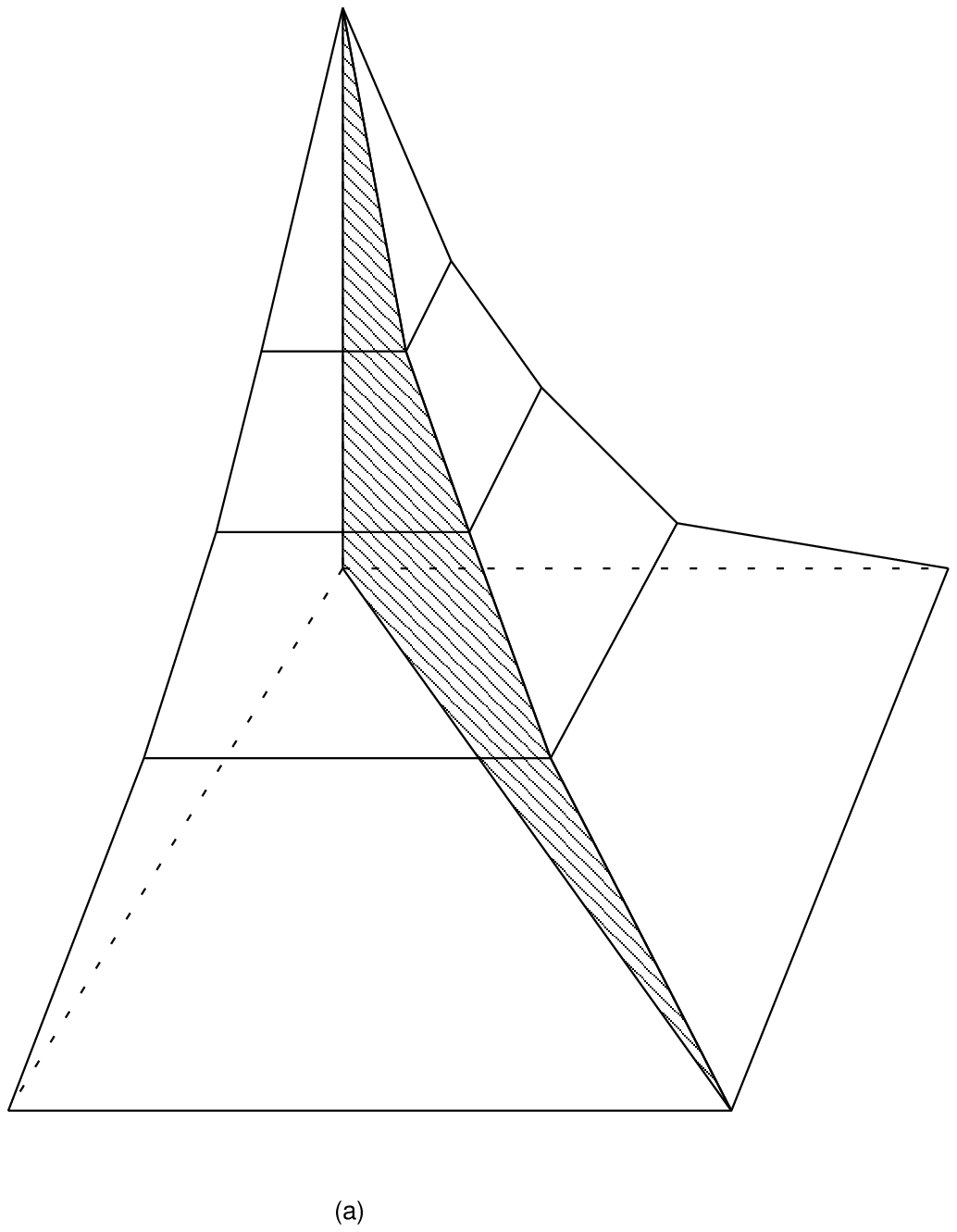}
    \end{minipage}
    &
    \begin{minipage}{70mm}
     \psfrag{(b)}{$(b)$}
     \psfrag{TN}{$T_N$}
     \includegraphics[scale=.6]{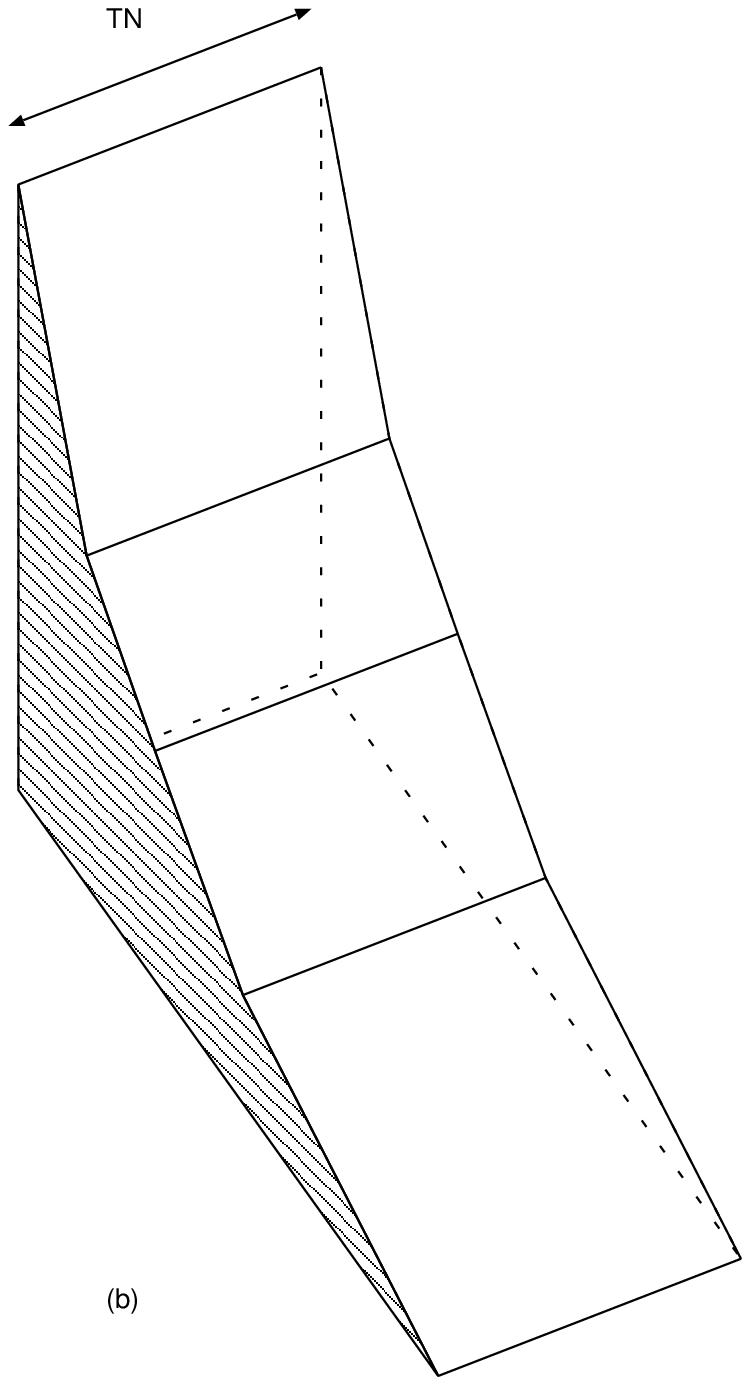}
    \end{minipage}
    \\
    \begin{minipage}{90mm}
     \psfrag{(c)}{$(c)$}
     \includegraphics[scale=.6]{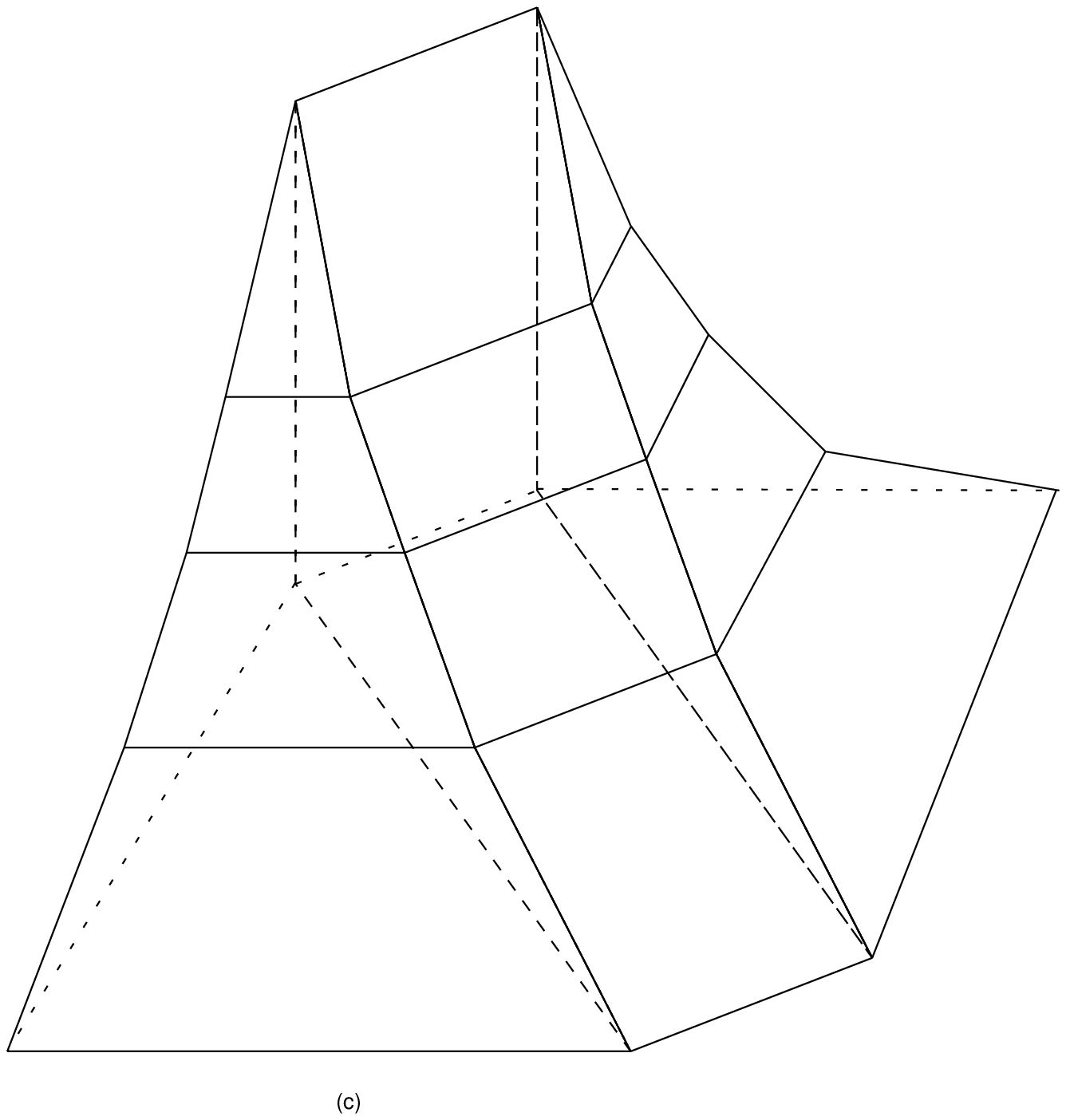}
    \end{minipage}
    &
   \end{tabular}
   \caption{\textit{Three-dimensional solids 
   obtained from the ground plane partition}}
   \label{fig;solidGPP}
  \end{center}
\end{figure}

Let us make a few comments. 
The first is about (\ref{BW by (q,T_N)}). 
This expression for the Boltzmann weight 
suggests that, in the hamiltonian approach   
the potential term $Q^{|\pi(0)|}$ 
might as well be absorbed as a jump of discrete time. 
Note that the transfer matrix in (\ref{eq;top_u(1)}) 
can be replaced with the following one.   
\begin{eqnarray}
 \prod\limits_{m=-\infty}^{-1}
 \Gamma_{+}(q^{-(m-\frac{T_N}{2}+\frac{1}{2})})
\cdot 
1^{T_N}
\cdot
 \prod\limits_{m=0}^{\infty} 
 \Gamma_{-}(q^{-(m+\frac{T_N}{2}+\frac{1}{2})})\,.
\end{eqnarray}
This implies that the potential term delays or 
advances the discrete time by the amount of $T_N$. 
The second is on (\ref{Boltzmann wt vs volume}). 
When the $U(1)$ charges are very large, 
we can approximate the ground plane partition, 
regarding as the three-dimensional Young diagram, 
to a three-dimensional solid.
See Figure \ref{fig;solidGPP}(a). 
The asymptotics of the ground partition 
is shaded in the figure. 
The size of ground plane partition in 
(\ref{Boltzmann wt vs volume}) is estimated as its volume. 
This consideration is generalized to the contribution of 
ground partition in (\ref{Boltzmann wt vs volume}), 
by thinking about another solid (Figure \ref{fig;solidGPP}(b)). 
By taking the first comment into account, 
we can imagine a new solid obtained from the two  
by cut and paste. Figure \ref{fig;solidGPP}(c).
Eq.(\ref{Boltzmann wt vs volume}) says that 
the volume of the last solid, 
measured by letting the unit cube in the plane partitions 
be one, is equal to 
$N$ times the volume of $\mathcal{P}^c(D)$, 
measured by letting the minimal volume 
in the K$\ddot{\mbox{a}}$hler gravity be one.

\subsubsection{Ground plane partition and quantum foam}

\begin{figure}[hbt]
 \psfrag{ex}{$e_x$}
 \psfrag{ey}{$e_y$}
 \begin{center}
 \includegraphics{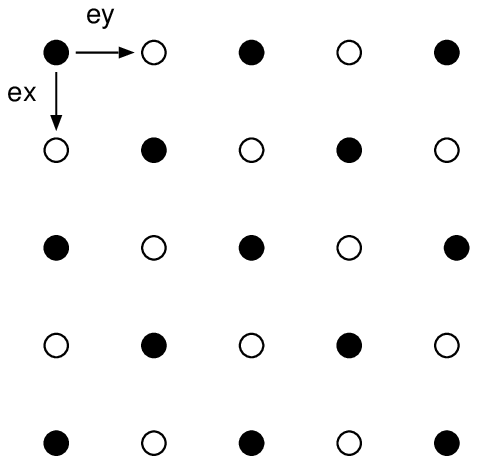}
 \caption{\textit{The sublattice $M^{\mbox{\tiny{PP}}}$ 
          consists of the bullets in the lattice $L$.}} 
 \label{fig;Lattice_L}
 \end{center}
\end{figure}

Three-dimensional Young diagrams are interpreted as  
the sets of lattice points, 
by regarding each cube as a lattice point. 
Let $\mbox{Supp}(\pi)$ denote the set of 
$(i,j)$ that satisfy $\pi_{ij}\neq 0$. 
A plane partition $\pi$ is identified with 
\begin{eqnarray}
\pi=
\Bigl\{ 
(i,j,k) \in \mathbb{Z}^{\oplus 3}\,;\,
(i,j) \in \mbox{Supp}(\pi),\,
1 \leq k \leq \pi_{ij} 
\Bigr\}\,. 
\label{PP as lattice points}
\end{eqnarray}
We will realize the lattice as a degree two sublattice of $L$ 
\begin{eqnarray}
M^{\mbox{\tiny{PP}}}
\equiv 
\mathbb{Z}(e_x-e_y)+\mathbb{Z}(e_x+e_y)
+\mathbb{Z}e_z\,, 
\label{M_PP}
\end{eqnarray}
by the mapping 
\begin{eqnarray}
(i,j,k) \,
\longmapsto \, 
\left\{ 
\begin{array}{cc}
(i-1)e_x+(2j-i-1)e_y+(k-1)e_z & \hspace{4mm} j-i \geq 1 \\
(i-1)e_x+(j-1)e_y+(k-1)e_z &  \hspace{4mm} j=i \\
(2i-j-1)e_x+(j-1)e_y+(k-1)e_z & \hspace{4mm} j-i \leq -1\,. 
\end{array}
\right. 
\label{map to M_PP}
\end{eqnarray}
See Figure \ref{fig;Lattice_L}.

We may divide the lattice points into the subsets 
$\pi^{(+)},\pi^{(-)}$ and $\pi^{(0)}$ 
by regarding plane partitions as the series of partitions.
They respectively consist of lattice points that come from 
the partitions $\pi(m)$ at $m \geq 1, m \leq -1$ and $m=0$. 
Let $\widehat{\pi}^{(+)}(T_N), \widehat{\pi}^{(-)}(T_N)$ 
and $\widehat{\pi}^{(0)}(T_N)$ denote sets of lattice points in $L$ 
that are determined from $\pi^{(+)},\pi^{(-)}$ and $\pi^{(0)}$ by 
\begin{eqnarray}
\widehat{\pi}^{(+)}(T_N)&=& 
\pi^{(+)}+T_Ne_y \,, 
\nonumber \\
\widehat{\pi}^{(-)}(T_N)&=& 
\pi^{(-)}+T_Ne_x \,, 
\label{parts for hat pi(T_N)}
\\
\widehat{\pi}^{(0)}(T_N)&=& 
\bigcup_{k=0}^{T_N}
\Bigl\{ 
\pi^{(0)}+T_Ne_x+k(-e_x+e_y) 
\Bigr\} \,. 
\nonumber
\end{eqnarray}
The union of the above, 
which we call $\widehat{\pi}(T_N)$, 
clearly incorporates the jumping effect. 
In particular, it satisfies  
$|\widehat{\pi}(T_N)|=|\pi|+T_N|\pi(0)|$. 
Note that translations involving $T_Ne_x$ and $T_Ne_y$ 
in (\ref{parts for hat pi(T_N)}) make  
$\widehat{\pi}(T_N) \subset M^{\mbox{\tiny{PP}}}$ 
or 
$\widehat{\pi}(T_N) \subset M^{\mbox{\tiny{PP}}}+e_x$. 

Let $\widehat{\pi}_{\mbox{\tiny{GPP}}}(p_r;T_N)$ be 
the set of lattice points in $L$ that is obtained from 
$\pi_{\mbox{\tiny{GPP}}}(p_r)$ as above. 
It can be observed that 
$\widehat{\pi}_{\mbox{\tiny{GPP}}}(p_r;T_N)$ 
becomes approximately 
$P_{(0,\cdots,0,T_N)}\setminus P_T$ 
when $T$ are very large. 
Actually we find  
\begin{eqnarray}
\widehat{\pi}_{\mbox{\tiny{GPP}}}(p_r;T_N) 
\cap \iota(M)
&=& 
\bigl( P_{(0,\cdots,0,T_N)}\setminus P_T \bigr)
\cap \iota(M) 
\nonumber \\
&=& 
\iota 
(\mathcal{P}^c(D) \cap M)\,. 
\label{PP vs P_c(D) cap M}
\end{eqnarray}
Note that $\iota(M)$ can be considered as a degree $N$ 
sublattice of $M^{\mbox{\tiny{PP}}}$. 
This implies that each gravitational quantum of the $SU(N)$ 
geometry consists of $N$ cubes of plane partitions.   
This also explains the factor $N$ which appears in 
(\ref{Boltzmann wt vs volume}).

\subsection{Generalized models of random plane partitions}

In order to relate random plane partitions 
with local $SU(N)$ geometries having other framings 
or five-dimensional gauge theories 
with the Chern-Simons terms taking other coupling constants, 
a slight modification of the model (\ref{Z(q,Q)}) is required. 
The relevant model is defined by 
\begin{eqnarray}
Z(q,Q\,;\,\widetilde{k}_{\mbox{\tiny{C.S.}}})
\equiv 
\sum\limits_{\pi} 
q^{|\pi|+\frac{\widetilde{k}_{\mbox{\tiny{C.S.}}}}{2N}\kappa(\pi(0))}\,
Q^{|\pi(0)|}\,. 
\label{eq;part_func_k}
\end{eqnarray} 
By repeating the previous argument 
and also taking account of (\ref{N-1 torsions}), 
one can find particularly  
\begin{eqnarray}
-\ln\, 
q^{|\pi_{\mbox{\tiny{GPP}}}(p_r)|
+\frac{\widetilde{k}_{\mbox{\tiny{C.S.}}}}{2N}
\kappa(\mu_{\mbox{\tiny{GP}}}(p_r))}\,
Q^{|\mu_{\mbox{\tiny{GP}}}(p_r)|}
=
\frac{1}{\hbar^2} 
\Bigl\{ 
\mathcal{F}^{pert}_{\mbox{{\scriptsize 5dSYM}}}
(a,\,\Lambda,\,R)
+O(\hbar) 
\Bigr\}\,, 
\end{eqnarray}
where the gauge theory prepotential is 
(\ref{perturbative prepotential with k}). 
This describes the case of the Chern-Simons term 
taking the coupling constant $k_{\mbox{\tiny{C.S.}}}$.

The transfer matrix for the above model naturally 
involves the charges of higher spin currents of 
$W_{1+\infty}$ algebra \cite{S-S} by 
(\ref{eq:L_0_abd_partition}). 
\begin{eqnarray}
Z(q,Q\,;\,\widetilde{k}_{\mbox{\tiny{C.S.}}})= 
\langle \emptyset;0 | 
\Bigl(
  \prod\limits_{m=-\infty}^{-1}
   \Gamma_{+}(q^{-(m+\frac{1}{2})})
\Bigr)\,
Q^{L_0}\,
q^{\frac{\widetilde{k}_{\mbox{\tiny{C.S.}}}}{2N}W^3_0}\,
\Bigl(
  \prod\limits_{m=0}^{\infty} 
   \Gamma_{-}(q^{-(m+\frac{1}{2})})
\Bigr)
|\emptyset ; 0\rangle\,,
\label{eq;corre_part_func_k}
\end{eqnarray}
where $L_0=W^2_0$ and $W^3_0$ are respectively 
the charges of the spin two and spin three currents 
(\ref{eq;eq_low_spin_op}). 
This seems to suggest some physical/geometrical importance 
in a further generalization of the statistical model such as 
\begin{eqnarray}
Z(q\,;\,s_2,\,s_3,\,\ldots)= 
\langle \emptyset;0 | 
\Bigl(
  \prod\limits_{m=-\infty}^{-1}
   \Gamma_{+}(q^{-(m+\frac{1}{2})})
\Bigr)\,
q^{\sum\limits_{n=2}^{\infty}s_nW_0^n}\,
\Bigl(
  \prod\limits_{m=0}^{\infty} 
   \Gamma_{-}(q^{-(m+\frac{1}{2})})
\Bigr)
|\emptyset ; 0\rangle\,. 
\label{generalized RPP}
\end{eqnarray}

\appendix
\newpage
\section{$2d$ Free Fermions, Partitions and $W_{1+\infty}$-Algebra} 
\label{sec;part_w}

Let 
$\psi(z)=\sum\limits_{r \in \mathbb{Z}+\frac{1}{2}}
\psi_rz^{-r-\frac{1}{2}}$ and 
$\psi^*(z)=\sum\limits_{s \in \mathbb{Z}+\frac{1}{2}}
\psi^*_sz^{-s-\frac{1}{2}}$ 
be complex fermions with the anti-commutation relations 
$\left\{\psi_r,\psi_s^*  \right\} = \delta_{r+s\,,\,0}$. 
In the fermion system, 
ground state for the charge $p$ sector 
is defined by the conditions 
\begin{eqnarray}
\psi_r |\emptyset ;p\rangle = 0 
&&
\mbox{for $r>-p$}\,, 
\nonumber \\
\psi_s^* |\emptyset ;p\rangle = 0 
&&
\mbox{for $s>p$}\,. 
\end{eqnarray}
For each partition $\mu$, 
the corresponding charged state 
$|\mu;\,p \rangle$ is built on the ground state 
as described in (\ref{eq;c_part_state}).

$W_{1+\infty}$ algebra is obtained 
\cite{S-S} from the algebra of psedo-differential operators 
by its central extension. 
Let us put $D=z\frac{d}{dz}$. Consider 
\begin{eqnarray}
W[z^n D^k] \equiv 
\oint \frac{dz}{2\pi i} :\psi(z)z^n D^k\psi^*(z):\,, 
\end{eqnarray}
where $:\hspace{2mm}:$ denotes 
the standard normal ordering and is prescribed by 
\begin{eqnarray}
:\,\psi(z)\psi^*(w): \hspace{2mm}
=\psi(z) \psi^*(w)-\frac{1}{z-w}\,.
\end{eqnarray}
These operators constitute the algebra. 
\begin{eqnarray}
\Bigl[  W[z^{n_1} D^{k_1}]\,,\, 
        W[z^{n_2} D^{k_2}] 
\Bigr]
&=&
W \bigl[
       z^{n_1+n_2} 
       \bigl\{
             (D+n_2)^{k_1} D^{k_2} -D^{k_1}(D+n_1)^{k_2} 
       \bigr\} 
   \bigr]
\nonumber \\   
&&
+\Psi (z^{n_1} D^{k_1},z^{n_2} D^{k_2}) \,\,,
\label{eq;W1+inf} 
\end{eqnarray}
where $\Psi$ denotes the central extension. 
It becomes possible \cite{w_al} to choose 
a basis for the algebra 
so that central charges between different 
spin operators vanish. In such a basis 
the lower spin operators have the following form. 
\begin{eqnarray}
W^1_n 
&=& 
W[z^n]\,, 
\nonumber \\
W^2_n 
&=& 
-W[z^n D]-\frac{n+1}{2}W[z^n]\,, 
\label{eq;eq_low_spin_op}\\
W^3_n 
&=& 
W[z^nD^2] +(n+1)W[z^nD] +\frac{(n+1)(n+2)}{6}W[z^n]\,. 
\nonumber
\end{eqnarray}
Among them, 
$J_n\equiv W^1_n$ and $L_n\equiv W^2_n$ constitute 
the Virasoro subalgebra. 
\begin{eqnarray}
\left[ J_k\,,\,J_{k'} \right]
&=& 
k \delta_{k+k',\,0}\,\,, 
\nonumber \\
\left[ L_m\,,\,J_k \right]
&=& 
-k J_{k+n} \,\,, 
\label{eq;vir} \\
\left[ L_m\,,\,L_n \right]
&=& 
(m-n)L_{m+n}
+\frac{n}{12}(n^2 - 1)\delta_{n+m\,,\,0}\,\,.  
\nonumber 
\end{eqnarray}

Charged partition states (\ref{eq;c_part_state}) 
become simultaneous eigenstates of infinitely many charges 
in $W_{1+\infty}$ algebra. 
\begin{eqnarray}
W 
\bigl[ e^{-t(D+\frac{1}{2})} \bigr]\,
|\mu ;\,p \rangle 
= 
\Bigl( 
P_{\mu,\,p}(t)-P_{\emptyset,\,p}(t)
+e^{\frac{t}{2}} \frac{e^{pt}-1}{e^t-1}  
\Bigr)
|\mu ;\,p \rangle\,,
\end{eqnarray}
where the operator in the LHS is a generating function 
of the charges, and $P_{\mu,\,p}(t)$ in the RHS denotes 
the characteristic function \cite{B-O} defined by  
\begin{eqnarray}
P_{\mu,\,p}(t)=
\sum\limits_{i=1}^{\infty}
e^{(p+\mu_i-i+\frac{1}{2})t}\,.
\label{characteristic function for charged partition}
\end{eqnarray}
For the lower spin operators (\ref{eq;eq_low_spin_op}), 
the above gives rise to  
\begin{eqnarray}
W^1_0\, |\mu;\,p \rangle 
&=& 
p\,|\mu;\,p \rangle\,, 
\nonumber \\
W^2_0\,|\mu;\,p \rangle 
&=& 
\Bigl( |\mu|+\frac{p^2}{2} \Bigr)
\,|\mu;\,p \rangle\,, 
\label{eq:L_0_abd_partition}\\
W^3_0\,|\mu;\,p \rangle 
&=& 
\Bigl(\kappa(\mu)+\frac{p^3}{3}\Bigr)
\,|\mu;\,p \rangle\,.
\nonumber 
\end{eqnarray}

\newpage
\subsection*{Acknowledgements}
We thank to T. Kimura for a useful discussion 
on toric geometries.
T.N. is supported in part by Grant-in-Aid for 
Scientific Research 15540273.



\begin{thebibliography}{99}



\bibitem{Crystal}
A.~Okounkov, N.~Reshetikhin and C.~Vafa,  
\textit{``Quantum Calabi-Yau and Classical Crystals,''} 
\texttt{hep-th/0309208}. 




\bibitem{quantum foam}
A.~Iqbal, N.~Nekrasov, A.~Okounkov and C.~Vafa, 
\textit{``Quantum Foam and Topological Strings,''}
\texttt{hep-th/0312022}. 




\bibitem{kahler gravity}
M.~Bershadsky and V.~Sadov, 
\textit{``Theory of K$\ddot{\mbox{a}}$hler Gravity,''}
Int. J. Mod.  Phys. \textbf{A11} (1196) 4689, 
\texttt{hep-th/9410011}


\bibitem{topological A_model}
E.~Witten, 
\textit{``Two-dimensional gravity and intersection theory on 
moduli space,''} 
Surveys Diff. Geom. \textbf{1} (1991) 243. 




\bibitem{topological vertex}
A.~Iqbal, 
\textit{``All Genus Topological String Amplitudes 
and 5-brane Webs as Feynman Diagrams,''}
\texttt{hep-th/0207114}. \\
M.~Aganagic, A.~Klemm, M.~Marino and C.~Vafa, 
\textit{``The Topological Vertex,''} 
\texttt{hep-th/0305132}.





\bibitem{Iqbal-Eguchi}
A. Iqbal and A.-K. Kashani-Poor, 
\textit{``$SU(N)$ geometries and topological string amplitudes,''}
\texttt{hep-th/0306032}. \\
T. Eguchi and H. Kanno, 
\textit{``Topological strings and Nekrasov's formulas,''}
JHEP \textbf{12} (2003) 006, 
\texttt{hep-th/0310235}. 






\bibitem{N-O}
N.~A.~Nekrasov,
\textit{``Seiberg-Witten Prepotential from Instanton Counting,''}
Adv. Theor.  Math.  Phys.  \textbf{7}  (2004) 831, 
\texttt{hep-th/0206161}. \\
N.~Nekrasov and A.~Okounkov,
\textit{``Seiberg-Witten Theory and Random Partitions,''}
\texttt{hep-th/0306238.}







\bibitem{MNTT}
T.~Maeda, T.~Nakatsu, K.~Takasaki and T.~Tamakoshi, 
\textit{``Five-Dimensional Supersymmetric Yang-Mills Theories 
and Random Plane Partitions,''}
JHEP \textbf{0503} (2005) 056, 
\texttt{hep-th/0412327}.






\bibitem{geometric engineering}
A.~Klemm, W.~Lerche, P.~Mayr, C.~Vafa and N.~P.~Warner,
\textit{``Self-Dual Strings and N=2 Supersymmetric Field Theory,''}
Nucl.  Phys. \textbf{B477} (1996) 746, 
\texttt{hep-th/9604034}. \\
S.~Katz, A.~Klemm and C.~Vafa,
\textit{``Geometric Engineering of Quantum Field Theories,''}
Nucl.  Phys.  \textbf{B497} (1997) 173, 
\texttt{hep-th/9609239}.




\bibitem{geometric quantization}
N.~Woodhause, 
\textit{``Geometric Quantization,''} 
Oxford  Univ. Press,  1992. \\
S.~Bates and A.~Weinstein, 
\textit{``Lectures on the Geometry of Quantization,''}
Berkeley Mathematics Lecture Notes \textbf{8}, 
AMS, 1991. 





\bibitem{Seiberg}
N.~Seiberg,
\textit{``Five Dimensional SUSY Field Theories, 
Non-Trivial Fixed Points and String Dynamics,''}
Phys.  Lett. \textbf{B388}  (1996) 753, 
\texttt{hep-th/9608111}.



\bibitem{triple intersection}
K.~Intriligator, D.~Morrison and N.~Seiberg,
\textit{``Five-Dimensional Supersymmetric Gauge Theories 
and Degenerations of Calabi-Yau Spaces,''}
Nucl.  Phys. \textbf{B497} (1997) 56, 
\texttt{hep-th/9702198}.\\*
A.~Iqbal and V.~Kaplunovsky,
\textit{``Quantum Deconstruction of a 5D SYM and its Moduli Space,''}
JHEP \textbf{0405}  (2004) 013, 
\texttt{hep-th/0212098}.





\bibitem{MNTT2}
T.~Maeda, T.~Nakatsu, K.~Takasaki and T.~Tamakoshi,
\textit{``Free fermion and Seiberg-Witten differential 
in random plane partitions,''}
Nucl.  Phys.  \textbf{B715} (2005) 275,
\texttt{hep-th/0412329}. 





\bibitem{Fulton}
W.~Fulton,
\textit{``Introduction to Toric Varieties,''}
Princeton University Press, 1993.



\bibitem{Oda}
T.~Oda,
\textit{``Convex Bodies and Algebraic Geometry,''}
Springer-Verlag, 1988.


\bibitem{Mirror}
C.~Vafa and E.~Zaslow eds, 
\textit{``Mirror Symmetry,''}
Clay Mathematics Monographs \textbf{1},
AMS, CMI,  2003.  




\bibitem{CS gravity}
E.~Witten, 
Nucl.  Phys. \textbf{B311} (1988/89) 46. 



\bibitem{coadjoint orbits}
J.~Navarro-Salas and P.~Navarro, 
\textit{``Virasoro Orbits, $\mbox{AdS}_3$ Quantum Gravity and Entropy,''}
JHEP \textbf{9905} (1999) 009, 
\texttt{hep-th/9903248},\\
T.~Nakatsu, H.~Umetsu, N.~Yokoi, 
\textit{``Three-Dimensional Black Holes and Liouville Field Theory,''}
Prog.  Theor.  Phys.  \textbf{102} (1999) 867, 
\texttt{hep-th/9903259}. 






\bibitem{Okounkov-Reshetikhin}
A.~Okounkov and N.~Reshetikhin,  
\textit{``Correlation Function of Schur Process with Application 
to Local Geometry of a Random 3-Dimensional Young Diagram,''} 
J. Amer. Math. Soc. \textbf{16} (2003) no.3 581, 
\texttt{math.CO/0107056}. 



\bibitem{Macdonald}
I.~G.~Macdonald,
\textit{``Symmetric Functions and Hall Polynomials,''}
Clarendon Press, 1995.




\bibitem{Miwa-Jimbo}
M.~Jimbo and T.Miwa,
\textit{``Solitons and Infinite Dimensional Lie Algebras,''}
Publ. RIMS, Kyoto Univ., \textbf{19} (1983) 943.





\bibitem{B-O}
S.~Bloch and A.~Okounkov, 
\textit{``The Character of the Infinite Wedge Representation,''}
\texttt{alg-geom/9712009}.



\bibitem{S-S}
M.~Sato and Y.~Sato, $in$ 
Nonlinear Partial Differential Equations in Applied Science; 
Proc. U.S.-Japan Seminar, Tokyo, 1982; 
Lect. Notes in Num.  Anal. \textbf{5} (1982) 259. 



\bibitem{w_al}
H.~Awata, M.~Fukuma, Y.~Matsuo and S.~Odake,
\textit{``Representation theory of the $W_{1+\infty}$ algebra,''} 
Prog.\ Theor.\ Phys.\ Suppl.  {\bf 118} (1995) 343, 
\texttt{hep-th/9408158}.






\end{thebibliography}
\end{document}